\documentclass[aps,prc,preprint,showpacs,preprintnumbers,nofootinbib,float,superscriptaddress,longbibliography]{revtex4-1}
\usepackage{graphicx, fancybox}
\usepackage{amsmath,amssymb}
\usepackage[colorlinks=true, pdfstartview=FitV, linkcolor=red,
		     citecolor=blue, urlcolor=blue]{hyperref}
\usepackage{color}
\usepackage{soul}
\usepackage{url}
\usepackage[normalem]{ulem}
\usepackage{nccmath}
\usepackage{amsthm}
\usepackage{mathrsfs}
\usepackage{graphicx}
\usepackage{caption}
\usepackage{subcaption}
\usepackage{float}

\usepackage{empheq}
\newcommand\widefbox[1]{\fbox{\rule[-1.8cm]{0pt}{5cm}\hspace{1em}#1\hspace{1em}}}
\newcommand\exactfbox[1]{\fbox{\hspace{1em}#1\hspace{1em}}}

\begin{document}
   
\title{\bf Torsional Newton Cartan gravity from non-relativistic strings}
\author{A.D. Gallegos, U. G\"ursoy and N. Zinnato}
\affiliation{Institute for Theoretical Physics, Utrecht University, Leuvenlaan 4, 3584 CE Utrecht, The Netherlands
\vspace{.5cm}}
\date{\today}


\begin{abstract}
\vspace{.3cm}
We study propagation of closed bosonic strings in torsional Newton-Cartan geometry based on a recently proposed Polyakov type action derived by dimensional reduction of the ordinary bosonic string along a null direction. We generalize the Polyakov action proposal to include matter, i.e. the   2-form and the 1-form that originates from the Kalb-Ramond field and the dilaton. We determine the conditions for Weyl invariance which we express as the beta-function equations on the worldsheet, in analogy with the usual case of strings propagating on a pseudo-Riemannian manifold. The critical dimension of the TNC space-time turns out to be 25. We find that Newton's law of gravitation follows from the requirement of quantum Weyl invariance in the absence of torsion. Presence of the 1-form requires torsion to be non vanishing. Torsion has interesting consequences, in particular it yields a mass term and an advection term in the generalized Newton's law. $U(1)$ mass invariance of the theory is  an important ingredient in deriving the beta functions. 
\end{abstract}

\maketitle

\newpage


\tableofcontents


\section{Introduction and Summary}\label{sec::intro}

Einstein's realization that gravity stems from geometrization of the Lorentz symmetry is among the greatest achievements in the history of physics. In general relativity, the equivalence principle is guaranteed by endowing spacetime with a (pseudo-)Riemannian structure that ensures the local Lorentz invariance. This profound connection between geometry and gravity is not unique to the laws of special relativity however, as an analogous connection exists also for the Galilean invariance. A covariant treatment of Galilean symmetry was first presented by Cartan \cite{Cartan1923, Cartan2, Friedrichs} leading to the discovery of the Newton-Cartan (NC) geometry as the underlying structure of classical Newtonian gravity. Subsequent work \cite{Inonu1952, Bargmann:1954gh, LevyLeblond:1967zz, LeblondBook1} clarified the algebra of  spacetime transformations  and its representation theory that underlies the NC geometry. In particular it was shown in \cite{Andringa:2010it} that the NC geometry follows from gauging the Bargmann algebra, the U(1) central extension of the algebra of Galilean boosts, translations and rotations. Finally, the structure of the Newton-Carton geometry has been extended to include torsion \cite{Niels2014, connection1}, and referred to as the ``torsional Newton-Cartan" (TNC) geometry\footnote{See \cite{connection1} for a discussion on necessity of including torsion in this theory.}. A crucial element in this geometry is the presence of the U(1) gauge symmetry that corresponds to the aforementioned central charge and physically related to the conservation of mass. Non-relativistic gravity has recently been studied in the context of non-relativistic effective actions \cite{Son}, non-relativistic holography \cite{Hartong2014}, post-Newtonian expansions of general relativity \cite{Dieter}, and more recently in the context of string theory \cite{Niels2018}. 

In this paper we ask the question whether the TNC geometry can be UV completed in a consistent theory of quantum gravity and take a few first steps in answering this question in the context of bosonic string theory\footnote{Eventually one may need superstrings to tame tachyonic instabilities but we expect this be a natural extension of the calculations we present here.}. 
One of the triumphs of the ordinary (relativistic) string theory has been the derivation of Einstein's equations in the weak gravity limit by demanding Weyl invariance of the world-sheet sigma model \cite{Callan}.   
In our case of string propagating on a manifold with local Galilean invariance, we similarly expect that the demand of quantum Weyl invariance on the world-sheet yields the Newton's law. This is what we mean precisely by the consistency of the TNC geometry with quantum gravity. 

Various proposals  to realize the Galilean symmetries in string theory exist in the literature. The Newton-Cartan geometry has only recently been  embedded  in string theory at the classical level, that is at the tree level of the world-sheet non-linear sigma model \cite{Harmark2014,Niels2017,Niels2018}. A parallel and separate line of work \cite{Andringa:2012uz,Bergshoeff:2018yvt, Bergshoeff:2018vfn, Bergshoeff:2017dqq} which started by the original paper of Gomis and Ooguri \cite{Gomis:2000bd}  realized  the Galilean symmetry in the context of closed string theory in a particular contraction limit, and, continued by the very recent paper \cite{Gomis:2019zyu} that asks the same question we ask here but in the context of the Gomis-Ooguri theory\footnote{In spite of the various connections between the Gomis-Ooguri approach \cite{Gomis:2000bd} and the TNC approach \cite{Harmark2014}, explained for instance in  \cite{Niels2018}, one should view these two approaches separately. In some sense the former is ``top-down" a and the latter ``bottom-up" approach to strings in Galilean invariant backgrounds.}.

We will follow the route taken by the papers \cite{Niels2017,Niels2018} where a Polyakov type action for a string propagating in the TNC geometry was constructed. Taking this Polyakov action as our starting point, we extend it to include bosonic target space matter, i.e. the Kalb-Ramond field $\bar B_{\mu\nu}$ and dilaton $\phi$, as well as an extra Kalb-Ramond 1-form $\aleph_\mu$, and we determine both the target space and worldsheet symmetries of this action at the classical level. We then go beyond the tree level and construct the worldsheet perturbation theory in powers of the string length $l_s$, assuming that the target TNC space is weakly curved. We then obtain the target space equations of motion from quantum Weyl invariance of the non-linear sigma model proposed in \cite{Niels2018} and its generalizations including the Kalb-Ramond and the dilaton fields. 

Here we summarise our main results that are the equations of motion that follow from the requirement of world-sheet Weyl invariance. Such equations are given in terms of the Galilean invariant TNC background fields $\{\tau_m, \bar h_{mn}, \hat \upsilon^m, \Phi \}$, the dilaton $\phi$, the Kalb-Ramond three form field strength $H= d \bar B$, the Kalb-Ramond two form field strength $\mathfrak{h}=d\aleph$, and the acceleration and electric fields $\{a_m,\mathfrak{e}_m\}$ defined via the twistlessness constraints $d \tau= a \wedge \tau$ and $d \aleph = \mathfrak{e} \wedge \tau$. In particular we find the equations of motion of the TNC target space as 

\begin{empheq}[box=\widefbox]{align*}
 h^{rs} D_ra_s + h^{rs}a_r a_s &= 2 h^{rs}\mathfrak{e}_r \mathfrak{e}_s + 2 h^{rs} a_r D_s \phi  \, ,\\
 h^{rs}D_r \mathfrak{e}_s &= 2 h^{rs} \mathfrak{e}_r D_s \phi \, ,\\
   R_{(mn)} - \frac{H_{rs(m} H_{n)tw} h^{rt} h^{sw}}{4} + 2 D_{(m} D_{n)} \phi &= h^{tq}\bar h_{q(m} D_{n)}a_{t} + \frac{a_m a_n}{2} + \mathfrak{e}_r h^{rs} \hat \upsilon^t \tau_{(m} H_{n)ts}   \\
   &\hphantom{=} +\frac{\mathfrak{e}^2 \left(2 \Phi \tau_m \tau_n - \bar h_{mn} \right) - \mathfrak{e}_m \mathfrak{e}_n}{2}     -a^2\, \Phi \tau_m\tau_n \, , \\
\frac{1}{2} h^{rs} D_r H_{smn} - h^{rp}H_{pmn}D_r\phi &= h^{tq}\bar h_{q[m} D_{n]} \mathfrak{e}_t +\hat \upsilon^r \tau_{[m} D_{n]} \mathfrak{e}_r  -a_{[m} \mathfrak{e}_{n]}\\
   &\hphantom{=}  - \frac{1}{2} a_r h^{rs}H_{smn}  + \left( \hat \upsilon^t D_t \phi - \frac{D_t \hat \upsilon^t}{2} \right)\mathfrak{h}_{mn} \, 
\end{empheq}
 where the covariant derivative $D$ and Riemann tensor $R_{mn}$ are defined with respect to the standard TNC connection 
 \begin{align}\label{TNCconnectionIntro}
\Gamma^m_{rs} \equiv - \hat \upsilon^m \partial_r \tau_s + \frac{1}{2} h^{m t} \left(\partial_r \bar h_{s t} + \partial_s \bar h_{r t} - \partial_t \bar h_{r s} \right)\, .
\end{align}

We follow the background field method to derive these equations. We start by splitting the embedding coordinates fields $X^m$ in a classical part $X^m_0$ and a quantum part $l_s \bar Y^m$ and use a covariant expansion of the TNC background fields to rewrite the action in the form of a perturbative series in quantum fluctuations parametrized by the string length $l_s$. A crucial step in this expansion is the construction of a set of normal coordinates $Y^m$ via a solution to the TNC geodesic equation \cite{Andringa:2012uz}
\begin{align}
	\ddot X^t + \mathring{\Gamma}^t_{mn} \dot X^m \dot X^n &= \frac{\dot N}{N}\dot X^t \, ,
\end{align}
with $N \equiv \tau_m \dot X^m$, and  boundary conditions $X^\mu(0)=X^m_0$ and $X^\mu(1) = X^m_0 + l_s \bar Y^m$. The normal coordinate $Y^m$ is defined as the tangent vector at the origin $\dot X^\mu(0) = l_s Y^m$. A new connection $\mathring{\Gamma}$ naturally arises 

 \begin{empheq}[box=\exactfbox]{align*}\label{TNCconnectionIntro}
	\mathring{\Gamma}^m_{rs} &= \Gamma^m_{rs} + \hat \upsilon^m \partial_{[r} \tau_{s]} + \frac{1}{2} a_n h^{mn} \bar h_{rs} \, .
\end{empheq}

Connection $\mathring{\Gamma}$ turns out to be invariant under all TNC symmetries unlike the standard connection $\Gamma$ which is non-invariant under the TNC U(1) mass symmetry for non-vanishing  torsion $d\tau \neq 0$. For the covariant expansion to exist a solution to the geodesic equation must be constructed. We find this to be possible only when the twistlessness constraint $d\tau=a \wedge \tau$ is satisfied. 

Our paper is organized as follows. We begin, in section \ref{PolyakovSection}, by reviewing the Polyakov-type action we used for the closed bosonic string moving in a TNC background and then generalize it to include the Neveu-Schwarz background matter, i.e. the dilaton and the Kalb-Ramond fields.  We then discuss how the target space and worldsheet symmetries are realized at the classical level. Section \ref{QWI} constitutes the core of our paper where we introduce the covariant background field expansion. This expansion coincides with the derivative expansion in the target space. We truncate this series at the second order both in the target-space derivatives and in the quantum fluctuations. Using this quantum effective action at the quadratic level, we then compute the one loop contribution to the Weyl anomaly and obtain the equations of motion for the TNC geometry arising from the vanishing of the beta functions. Finally in section \ref{Conclusions} we present a discussion of the results and provide an outlook. Several appendices where we give details of our (quite lengthy) calculations form a substantial part of this paper. 

{\it Notes addded:} We became aware of a paper of Gomis, Oh and Yan \cite{Gomis:2019zyu} on the quantum Weyl symmetry of the non-linear sigma model for the non-critical string theory in the final stage of our work. \\ {\it v2:} The second version of the paper contains substantial improvements over the first one. In particular here we use the aforementioned geodesic equation to define the covariant expansion, which is consistent with the U(1) mass symmetry. 

\section{The string action and its symmetries}
\label{PolyakovSection}

\subsection{The Polyakov action without matter} 

The geometric data of the TNC geometry in the absence of matter fields is encoded in a pair of vielbeins\footnote{We will use letters $\{m,n,... \}$ to denote curved TNC indices and $\{i,j,...\}$ to denote flat TNC indices.} $(\tau_s, e^i_s)$ and a U(1) connection $m_s$ collectively referred as the TNC metric complex. The vielbeins $e^i_s$ define a degenerate spatial metric through $h_{m n}= e_m^i e_n^j \delta_{i j}$  and it is possible to use the inverse of the square matrix $(\tau_m, e^i_n )$, denoted as  $(-  \upsilon^m, e^n_i)$  with $ \upsilon^m \tau_m = -1$ and  $\tau_m e^m_i=0$, to define an independent spatial inverse metric $h^{m n}= e^m_i e^n _j \delta^{i j}$. These spatial metrics together with the temporal coframes, $\tau_m$ and $\upsilon^m$, are subject to a completeness relation $\delta^m_n = -\upsilon^m \tau_n + h^{m r} h_{r n} $. 

\vspace{2mm}
Quite conveniently, the TNC geometry with this geometric data can be derived from a higher dimensional relativistic spacetime with an isometry in the extra null direction---which we will denote as the $u$-direction---via the procedure of null reduction \cite{nullReduction}. In particular we consider the TNC manifold to be d+1-dimensional and the relativistic one will be d+2-dimensional. The metric of such relativistic spacetimes can always be written as
\begin{align}\label{uplift}
\bar g_{M N} dx^M dx^N = 2 \tau \left(du-m \right) + h_{m n} dx^m dx^n, 
\end{align}
with $\partial_u$ the corresponding null Killing vector. We label indices of the d+2 dimensional space as $M=\{u,m\}$. We also define $\tau=\tau_m dx^m$, $m=m_s dx^s$ with $x^m$ the coordinates of the (d+1)-TNC manifold. It is now possible to derive the world-sheet action for a string moving in the TNC geometry \cite{Niels2017,Niels2018} starting from the ordinary Polyakov action  in the relativistic target space \eqref{uplift}: 
\vspace{2mm}
\begin{align}\label{PolyakovUplift}
\mathcal{L}= - \frac{\sqrt{-\gamma}}{4 \pi l_s^2}  \gamma^{\alpha \beta} \left(h_{\alpha \beta}- \tau_r m_s - m_r \tau_s\right) - \frac{\sqrt{-\gamma}}{2 \pi l_s^2} \gamma^{\alpha \beta} \tau_\alpha \partial_\beta X^u,
\end{align}
where $\gamma$ is the determinant of the worldsheet metric $\gamma_{\alpha \beta}$, and where $ h_{\alpha \beta} = \bar h_{r s} \partial_\alpha X^r \partial_\beta X^s$ and $\tau_\alpha= \tau_m \partial_\alpha X^m$ are the pullbacks of $ h_{r s}$ and $\tau_r$ respectively\footnote{We use the first few greek and latin indices $\{\alpha,\beta... \}$ and  $\{a,b,...\}$ to denote the curved and flat worldsheet indices respectively.}.

We consider a  closed string without winding, i.e. $X^m(\sigma^0,\sigma^1 + 2 \pi ) = X^m (\sigma^0,\sigma^1)$ , and with non zero momentum $P$ along $X^u$ 
\begin{align}
P= \int^{2\pi}_0 d \sigma^1 P^0_u\, , 
\end{align}
with the momentum current 
\begin{align}\label{Pu}
P^\alpha_u = \frac{\partial \mathcal{L}}{\partial \partial_\alpha X^u} &= -\frac{\sqrt{-\gamma} \gamma^{\alpha \beta} \tau_\beta}{2 \pi l_s^2}\, .
\end{align}
Following \cite{Niels2018} it is possible to rewrite \eqref{PolyakovUplift} in a dual formulation where the conservation of the momentum current \eqref{Pu} is implemented off-shell through the classically equivalent Lagrangian 
\begin{align}
\begin{split}\label{dualLagrangian}
\mathcal{L} =-\frac{\sqrt{-\gamma} \gamma^{\alpha \beta}\bar h_{\alpha \beta}}{4\pi l_s^2} - \frac{1}{2 \pi l_s^2} \left(\sqrt{-\gamma} \gamma^{\alpha \beta} \tau_\beta - \epsilon^{\alpha \beta} \partial_\alpha \eta \right) A_\beta \, ,
\end{split}
\end{align}
where $A_\alpha$ is a Lagrange multiplier that enforces conservation of $P^\alpha_u = \frac{\epsilon^{\alpha \beta} \partial_\beta \eta}{2\pi l_s^2}$ off-shell and we defined the combination
\begin{equation}
\label{hbar}
\bar h_{\alpha \beta} \equiv h_{\alpha \beta} - \tau_\alpha m_\beta - m_\alpha \tau_\beta \, .
\end{equation}
The significance of this combination will become clear when we discuss the symmetries of the theory below. 

This procedure introduces a novel degree of freedom, a scalar field $\eta$ on the world sheet. To see that \eqref{dualLagrangian} and \eqref{PolyakovUplift} are equivalent one uses the equation of motion for $\eta$ which gives $A_\alpha = \partial_\alpha \chi$ for some world sheet scalar  $\chi$ and identifies the latter with the $u$-direction $\chi=X^u$ recovering the original Lagrangian \eqref{PolyakovUplift}. Following \cite{Niels2018} we introduce the worldsheet zweibein $e^a_\alpha$ and its inverse $e^\alpha_a=\epsilon^{\alpha \beta} e^b_\beta \epsilon_{ba}$, satisfying $e^a_\alpha e^b_\beta \eta_{ab} = \gamma_{\alpha \beta}$ and $e^\alpha_a e^\beta_b \eta_{a b} = \gamma^{\alpha \beta}$, to rewrite the constraints as 
\begin{align}
\label{constraints}
\begin{split}
\epsilon^{\alpha \beta}\left( e^0_\alpha + e^1_\alpha \right) \left( \tau_\beta + \partial_\beta \eta \right) &= 0\, , \\
\epsilon^{\alpha \beta} \left(e^0_\alpha - e^1_\alpha \right) \left(\tau_\beta - \partial_\beta \eta \right) &= 0\, .
\end{split}
\end{align}
A final field redefinition 
\begin{align}
A_\alpha &= m_\alpha + \frac{1}{2} \left( \lambda_+ - \lambda_- \right) e^0_\alpha + \frac{1}{2} \left( \lambda_+ + \lambda_- \right) e^1_\alpha  
\end{align}
yields the Lagrangian 
\begin{align}
\label{proposal}
\mathcal{L}&= - \frac{1}{4 \pi l_s^2} \left[ 2 \epsilon^{\alpha \beta} m_\alpha \partial_\beta \eta + e \eta^{ab} e^\alpha_a e^\beta_b h_{\alpha \beta} - \lambda_+ e^\beta_- \left(\partial_\beta \eta + \tau_\beta  \right) - \lambda_- \left(\partial_\beta \eta - \tau_\beta \right)  \right],
\end{align}
where $e^\alpha_{\pm}= e^\alpha_0 \pm e^\alpha_1$. This is the Polyakov-type Lagrangian for a string moving in a TNC geometry proposed in  \cite{Niels2018}.  We further use the constraints to rewrite \eqref{proposal} in a way more convenient for quantization\footnote{One should think of implementing these constraints inside the Polyakov path integral to ensure equivalence of the quantum path integrals based on the lagrangians (\ref{proposal}) and (\ref{Polyakov}).}
\begin{align}
\begin{split}\label{Polyakov}
\mathcal{L} = \frac{e}{4 \pi l_s^2}  \left[ e^\alpha_+ e^\beta_- \bar h_{\alpha \beta} +  \lambda_+ e^\beta_- \left( \partial_\beta \eta + \tau_\beta \right) + \lambda_- e^\beta_+ \left(\partial_\beta \eta -\tau_\beta  \right)   \right],
\end{split}
\end{align}
We will examine the quantum path integral defined by this Lagrangian in the rest of the paper, but we will first extend it to include Neveu-Schwarz matter, i.e. the Kalb-Ramond field and dilaton and then discuss the symmetries of this generalized action both on the worldsheet and in the target space. 

\subsection{The Polyakov action with matter} 
\label{PolyakovSection2}  

It is straightforward to generalize the action (\ref{Polyakov})  to include standard Neveu-Schwarz matter, i.e. a Kalb-Ramond field $ \mathcal{B}_{MN}$ and a dilaton $\phi$. Let us first consider the B-field. 
Once again, to derive the corresponding Lagrangian we can start from its null lifted version. We then obtain the following action by rearranging the terms that follow from the null reduction of the relativistic d+2 dimensional bosonic Polyakov action with the B-field: 
\begin{align}
\begin{split}\label{classical1}
\mathcal{L}&= - \frac{1}{4 \pi l_s^2} \left(\sqrt{-\gamma} \gamma^{\alpha \beta} \bar h_{\alpha \beta} + \epsilon^{\alpha \beta} \bar B_{\alpha \beta}  \right) - \frac{1}{4 \pi l_s^2} \left(\sqrt{-\gamma} \gamma^{\alpha \beta} \tau_\alpha - \epsilon^{\alpha \beta} \aleph_\alpha \right) \partial_\beta X^u
\end{split}
\end{align}
where we defined 
\begin{align}
\label{aleph}
\aleph_\alpha &\equiv \mathcal{B}_{u \alpha}=-\mathcal{B}_{\alpha u}\, ,\\
\label{Bbar}
\bar B_{\alpha \beta} &\equiv \mathcal{B}_{\alpha \beta}  \, .
\end{align}
Following the same procedure as in \cite{Niels2018} described in section \ref{PolyakovSection} we compute the momentum along $X^u$
\begin{align}
P^\alpha_u &= - \frac{1}{2 \pi l_s^2} \left(\sqrt{-\gamma} \gamma^{ \beta\alpha} \tau_\beta - \epsilon^{\beta\alpha} \aleph_\beta  \right)
\end{align}
and implement its conservation off-shell via 
\begin{align}
\begin{split}\label{classical2}
\mathcal{L} &= - \frac{1}{4 \pi l_s^2} \left( \sqrt{-\gamma} \gamma^{\alpha \beta} \bar h_{\alpha \beta} + \epsilon^{\alpha \beta} \bar B_{\alpha \beta} \right) - \frac{1}{2 \pi l_s^2} \left( \sqrt{-\gamma} \gamma^{\alpha \beta} \tau_\alpha - \epsilon^{\alpha \beta} \aleph_\alpha - \epsilon^{\alpha \beta} \partial_\alpha \eta \right) A_\beta. 
\end{split}
\end{align}
Making, once again, the field redefinition 
\begin{align}
A_\alpha= m_\alpha + \frac{1}{2} \left(\lambda_+ - \lambda_- \right) e^0_\alpha + \frac{1}{2} \left(\lambda_+ + \lambda_- \right)e^1_\alpha\, , 
\end{align} 
integration over the worldsheet fields $\lambda_\pm$ now impose the constraints 
\begin{align}
\label{constraintsB}
\begin{split}
\epsilon^{\alpha \beta}\left( e^0_\alpha + e^1_\alpha \right) \left( \tau_\beta + \aleph_\beta+\partial_\beta \eta \right) &= 0\,,  \\
\epsilon^{\alpha \beta} \left(e^0_\alpha - e^1_\alpha \right) \left(\tau_\beta -  \aleph_\beta - \partial_\beta \eta \right) &= 0\, .
\end{split}
\end{align}
we can cast \eqref{classical2} in the Polyakov form  
\begin{align}
\begin{split}\label{PolyakovB}
\mathcal{L} &=  \frac{1}{4 \pi l_s^2} e \left[ e^\alpha_+ e^\beta_- \left(\bar h_{\alpha \beta} + \bar B_{\alpha \beta} \right) + \lambda_+ e^\beta_- \left( \partial_\beta \eta + \aleph_\beta + \tau_\beta\right) + \lambda_- e^\beta_+ \left(\partial_\beta \eta + \aleph_\beta - \tau_\beta  \right) \right],
\end{split}
\end{align}
where just as in \eqref{Polyakov} the constraints (\ref{constraintsB})  have been used. Lagrangian \eqref{PolyakovB} is still invariant under \eqref{WeylLorentz} and the contribution of the B-field to the anomaly can in principle be computed in a similar manner as performed for \eqref{Polyakov}. 

When the world-sheet is non-flat, in addition to the B-field, it is also possible to include a dilaton contribution of the form 
\begin{align}\label{dilac}
\mathcal{L}_\phi &= \frac{1}{16\pi} \sqrt{-\gamma} \mathcal{R} \phi \, ,
\end{align}
where $ \mathcal{R}$ is the worldsheet Ricci scalar. The Polyakov path integral then  involves a sum over world-sheet topologies that is organized in powers of $\exp(\phi)$ as usual. 
\vspace{-.4cm}
\subsection{Symmetries of the Polyakov action} 

We will now discuss both the target space and  world-sheet symmetries of the world-sheet action (\ref{PolyakovB}) and (\ref{dilac}).  

\subsubsection{Space-time symmetries} 
\label{syms}

The fields in the TNC metric complex, without matter, transform under diffeomorphisms $\xi$, local Galilean boosts  $\lambda^i$, local rotations $\lambda^{ij}$ and local U(1) gauge transformation $\sigma$ and the Lagrangian (\ref{dualLagrangian}) is invariant under these transformations \cite{Niels2018}. These transformations are easily generalized in the presence of matter. All in all, the transformations of all the objects that enter the calculations read
\begin{align}
\begin{split}\label{transformationLaws}
\delta \tau_s &= \pounds_\xi \tau_s, \\
\delta e^i_s &= \pounds_\xi e^i_s + \lambda^i \tau_s + \lambda^{i}_{\hphantom{i}j} e^j_s, \\
\delta \upsilon^s &= \lambda^i e^s_i, \\
\delta e^s_i &= \pounds_\xi e^s_i, \\
\delta m_s &= \pounds_\xi m_s + \lambda_i e^i_s + \partial_s \sigma, \\
\delta \bar B_{mn} &= \pounds_\xi \bar B_{mn} +2\aleph_{[m}\partial_{n]}\sigma\, ,\\
\delta \aleph_m &= \pounds_\xi \aleph_m\, ,\\
\delta \phi &= \pounds_\xi \phi\, .
\end{split}
\end{align}
In particular, the combination $\bar h_{mn}$ defined in (\ref{hbar}) and (\ref{Bbar}) is invariant under local Galilean boosts and transforms under local  $U(1)$ mass transformations as    

\begin{align}\label{hbTransf}
	\delta_\sigma \bar h_{mn} = -2 \tau_{(m} \partial_{n)}\sigma 
\end{align}

Now, it is straightforward to check that the actions based on (\ref{PolyakovB}) and  (\ref{dilac}) are invariant under diffeomorphisms, local Gallilean boosts, local rotations and local U(1) transformations. When starting with the explicitly Galilean boost invariant form \eqref{PolyakovB} it is crucial to use the constraints (\ref{constraintsB}) to show invariance under local $U(1)$ mass transformations.  However, the classical equations of motion will not be invariant under this $U(1)$ symmetry, see \eqref{boxfin}. To fix this we will ask that the Lagrange multipliers transform under the symmetry as
\begin{align} \label{lambdaU1}
\delta \lambda_+ = -e^\alpha_+\partial_\alpha \sigma\, ,\qquad\qquad\qquad  \delta \lambda_- = e^\alpha_-\partial_\alpha \sigma.
\end{align}
Taking this into account, both the action and the equations of motion can be shown to be $U(1)$ mass invariant off-shell. In what follows, in addition to $\bar h_{mn}$ and $\bar B_{mn}$ defined in (\ref{hbar}) and (\ref{Bbar}),  it will prove useful to introduce the following combinations 
\begin{eqnarray}
\label{barups}
\hat \upsilon^m &\equiv& \upsilon^m - h^{m s} m_s\, , \\
\label{Phi}
\Phi &\equiv& - \upsilon^s m_s + \frac{1}{2} h^{rs}m_r m_s \, ,
\end{eqnarray}
that are invariant under local Galilean boost and rotations as one can easily check using (\ref{transformationLaws}). They do transform under a local $U(1)$ mass transformation: 
\begin{align}
\label{trbarups}
\delta_\sigma \hat \upsilon^m &= - h^{mn}\partial_m\sigma\, \\
\label{trPhi}
\delta_\sigma \Phi & =-\hat \upsilon^n\partial_n \sigma\,  .
\end{align} 
 Even though they do not appear in the action at the classical level, we have introduced $\hat \upsilon^m$ as the local Galilean boost and rotations invariant version of $\upsilon^m$ the inverse of $\tau_m$, and the target space scalar $\Phi$ which will play the role of the Newton's gravitational potential below. They will become important when we discuss quantum corrections in the theory. We note that $\hat \upsilon^m$, $\tau_m$, $\bar h_{mn}$ and $h^{mn}$ are subject to the completeness relation $\delta^r_s = -\hat \upsilon^r \tau_s +  h^{r m} \bar h_{m s}$. 
Finally, we note that because of the non-trivial U(1) mass transformation of $\bar B_{mn}$ in 
(\ref{transformationLaws}), i.e. $\delta_\sigma \bar B = \aleph \wedge d\sigma$, the field strength, $H = d\bar B$ will transform under mass $U(1)$ as 

\begin{align}\label{HTransf}
	\delta_\sigma H_{mnp} &= \mathfrak{h}_{mn} \partial_p \sigma + \mathfrak{h}_{np} \partial_m \sigma + \mathfrak{h}_{p m} \partial_n \sigma \, ,
\end{align}
with 
\begin{equation}\label{defh}
\mathfrak{h}_{mn} \equiv \partial_m \aleph_n - \partial_n \aleph_m\, 
\end{equation} 
being the field strength of $\aleph$. Notice in particular that setting $H_{mnp}=0$ would not be a $U(1)$ mass invariant condition unless $\mathfrak{h}_{mn}=0$.
\subsubsection{$U(1)_B$ one-form symmetry} 
\label{oneformsym}

In the presence of the Kalb-Ramond field there is also a U(1) one-form symmetry. It is well-known that the transformation
\begin{equation}
\delta_\Lambda \mathcal{B}_{MN} = \partial_M \Lambda_N - \partial_N \Lambda_M ,
\end{equation} 
where $\partial_M$ is the partial derivative in the target space, is a symmetry of the d+2 dimensional world-sheet action with the relativistic target space. 

After null reduction the resulting TNC geometry with Kalb-Ramond matter has a U(1) one-form symmetry of the form: 
\begin{align}
\label{BLambda}
\delta_\Lambda \bar B_{mn} &= \partial_m \Lambda_n - \partial_n \Lambda_m  ,\\
\label{alephLambda}
\delta_\Lambda \aleph_m &= \partial_m \Lambda_u \, .
\end{align}

We see that in the TNC geometry $\aleph$ acquires a new local U(1) symmetry, whereas $B$ transforms under a local one-form symmetry. It is now straightforward to check that the action (\ref{PolyakovB}) is invariant under (\ref{BLambda}) upon use of the constraint equations (\ref{constraintsB}). Invariance of (\ref{PolyakovB}) under (\ref{alephLambda}) however requires a non-trivial transformation of the worldsheet field $\eta$: 
\begin{align}
\label{etaLambda}
\delta_\Lambda \eta = - \Lambda_u \, ,
\end{align}
which is a trivial shift in the quantum path integral where $\eta$ is path integrated. Therefore, we conclude that the action, at least at the tree-level, is invariant under both the local one-form symmetry $\Lambda_m$ and the new local U(1) symmetry $\Lambda_u$. 
The fact that $\eta$ is charged under the U(1) that comes from the B-field, i.e. eq. (\ref{etaLambda}), is expected as one can think of $\eta$ as the direction dual to $u$, \cite{Niels2018}. In this sense the gauge fields $m$ and $\aleph$ can be considered as dual to each other. 

In passing, we note that for $\aleph=0$ the action (\ref{PolyakovB}) enjoys an additional symmetry for $\bar B_{mn}$ given by 
\begin{align}\label{largeU(1)}
\delta \bar B_{m n} &= 2 \Omega(X) \, \partial_{[m} \tau_{n]} \, ,  
\end{align}
with $\Omega$ an arbitrary spacetime function satisfying $\partial_\alpha \partial^\alpha \Omega(X) = 0$. To show that \eqref{largeU(1)} is a symmetry it is necessary to use the constraint equations \eqref{constraintsB}. 

\subsubsection{Local worldsheet symmetries} 

The actions \eqref{Polyakov} and \eqref{PolyakovB} are clearly invariant under the worldsheet diffeomorphisms. These symmetries allow us to cast the worldsheet metric in a diagonal form $\gamma^{a b} = e^{-2 \rho} \eta^{ab}$ where the conformal factor $\rho$  determines the Ricci curvature of the worldsheet $\mathcal{R}$ (locally) as
\begin{align}
\sqrt{-\gamma} \mathcal{R}=  -2 \partial^2 \rho  \, .
\end{align}
We will refer to this choice of gauge as the conformal gauge. The reparametrization gauge-fixed  Polyakov Lagrangians \eqref{Polyakov} and \eqref{PolyakovB} further exhibit a residual Lorentz/Weyl gauge invariance of the form (as can be checked straightforwardly) 
\begin{align}
\begin{split}\label{WeylLorentz}
e^\alpha_{\pm} \rightarrow f_{\pm} e^a_{\pm} , \quad \quad \lambda_{\pm} \rightarrow f_{\pm} \lambda_{\pm},
\end{split}
\end{align}
for any worldsheet function $f_{\pm}$. For $f_+=f_-$ the transformation is a local Weyl transformation and for $f_+=-f_-$ it constitutes a local Lorentz transformation. Once we have used diffeomorphism invariance to go to conformal gauge it is possible to use local Weyl invariance to fix the mode $\rho$ and completely fix the worldsheet metric $\gamma^{\alpha \beta}$. 

The main purpose of our paper is to discuss the fate of these residual gauge invariances at the quantum level. Here it suffices to note that, in the case without matter, the condition for invariance of the Polyakov action $S(e,\lambda,X)$ under the gauge transformations \eqref{WeylLorentz} at the classical  level takes the form  
\begin{align}
\begin{split}\label{traceless}
\frac{\delta S}{\delta f_{\pm}}= e^\gamma_c \tau^c_\gamma +  C^+ \lambda_+ +  C^- \lambda_-  &=0,
\end{split}
\end{align}
where the energy momentum one form\footnote{Even though it is possible to define an energy momentum tensor from $\tau^c_\gamma$ via $T_{\alpha \beta}=\eta_{c d} e^d_\alpha \tau^c_\beta$ it is more natural to define the traceless condition in terms of the energy momentum one form.} $\tau^c_\gamma$ and constraint functions $C^{\pm}$ are defined as

\vspace{2mm}

\begin{align}
\tau^c_\gamma &\equiv- \frac{2 \pi l_s^2}{e} \frac{\delta S}{\delta e^\gamma_c} \nonumber \\
\label{momentumOne}
&= \frac{2 \pi l_s^2 \mathcal{L}}{e} e^c_\gamma + \frac{1 }{2} \left[ 2 e^\beta_b \eta^{c b} \bar h_{\gamma \beta} - \lambda_+ \left(\delta^c_0 - \delta^c_1 \right) \left( \partial_\gamma \eta + \tau_\gamma \right) - \lambda_- \left(\delta^c_0 + \delta^c_1 \right) \left( \partial_\gamma \eta - \tau_\gamma \right)\right]\, ,\\
\label{constraintsOne}
C^{\pm} &\equiv  - \frac{2 \pi l_s^2}{e} \frac{\delta S}{\delta \lambda_{\pm}} = -\frac{1}{2} e^\beta_{\mp} \left(\partial_\beta \eta \pm \tau_\beta \right).
\end{align}

\vspace{2mm}
The condition \eqref{traceless} is nothing but a constrained traceless condition for the energy momentum tensor, and from \eqref{momentumOne} and \eqref{constraintsOne} it is clear that this conditions holds for the Polyakov action \eqref{Polyakov}. The rest of our work will concern the computation of \eqref{traceless} at the quantum level, in particular, at the one-loop level in the perturbative expansion in $l_s^2$.

\section{Quantum weyl invariance of string in the TNC geometry }
\label{QWI}

\subsection{Background field quantization}
\label{backgroundExpansion}

The quantum partition function that follows from the action \eqref{Polyakov} is defined by the Polyakov path integral\footnote{It is crucial to include the contribution from the Faddeev-Popov ghosts that come from the gauge fixing but we will not explicitly show them here. The gauge fixing procedure is discussed in Appendix \ref{ghosts}.}. As for the bosonic strings \cite{Callan}, it will be very helpful to introduce the background field formalism to organize  the perturbative $l_s^2$ expansion to study the quantum properties of the worldsheet sigma model.  To this end, we expand the worldsheet fields $\{X^m, \lambda_{\pm},\eta \}$ around a classical configuration $\Psi_0\equiv\{X^m_0, \lambda^0_{\pm},\eta_0 \}$ as 
\begin{align}
\begin{split}\label{quantumFields}
X^m &= X^m_0 + l_s \bar Y^m, \\
\lambda_{\pm} &= \lambda^0_{\pm} + l_s \bar \Lambda_{\pm}, \\
\eta &= \eta_0 + l_s \bar H , 
\end{split}
\end{align}
where $\Psi\equiv\{\bar Y^m,\bar \Lambda_{\pm},\bar H\}$ below will collectively denote the quantum fields. Using this expansion, the one loop effective effective action $\Gamma[\Psi_0]$ for the background fields can be expressed \cite{YellowBook} as a path integral over the quantum fields as 

%
%
%
%
\begin{align}\label{oneLoopAction}
e^{i \bar \Gamma[\Psi_0](0)} &= \int D \Psi \hphantom{\hspace{1mm}} e^{i \bar S[\Psi_0,\Psi](0)}.
\end{align}
where $\bar S[\Psi_0,\Psi](0)$ is the $\mathcal{O}\left(l^0_s\right)$ term that arises from substituting (\ref{quantumFields})  in \eqref{Polyakov}. In \eqref{oneLoopAction} the zweibeins are completely fixed by the Faddeev-Popov procedure, see Appendix \ref{ghosts}, using the reparametrization invariance and Weyl symmetry. This, in particular,  fixes the function $\rho$. If the symmetry \eqref{WeylLorentz} is to be consistent at the one loop level then any change of $\rho$ should leave the effective action invariant, this means that the Weyl invariance \eqref{traceless} at the one loop level becomes \footnote{We are assuming that a path integral measure invariant under the target spacetime symmetries exists.}
\begin{align}\label{variationEffective}
\delta_\psi \bar \Gamma \left[ \Psi_0 \right] (0) =0, \quad \quad \delta_\psi \rho = \psi  
\end{align}
\subsection{Covariant background expansion}

The goal of this section is to express  $\bar S[\Psi_0,\Psi] (0)$ as an action over TNC covariant fields, for this we first note that $\bar Y^m$ does not transform as a vector under general coordinate transformations. To get covariant expressions we first need to rewrite $\bar Y^m$ covariantly.  This is achieved \cite{YellowBook} by considering a geodesic connecting $X_0^m$ and $X^m_0 + \bar Y^m$ to rewrite $\bar Y^m$ as  
\vspace{2mm}
\begin{align}
\begin{split}\label{RNC}
\bar Y^m = Y^m - \frac{l_s}{2} \left( \Gamma^m_{r s} +  G^m_{rs} \right)_0 Y^r Y^s + \mathcal{O}\left( l_s^2 \right),
\end{split}
\end{align}
where $Y^m$ is the tangent vector along the geodesic, $()_0$ indicates the corresponding expression is evaluated at $X_0$, $\Gamma^m_{rs}$ is the TNC connection characterising the non-covariant part of $\bar Y^m$, and $G^m_{rs}$ is a tensor symmetric in its lower indices and the solution to the tensor equation \footnote{A solution to \eqref{GEquationM} exists as long as the torsion is taken to be twistless, namely as long as $F_{mn} h^{mt} h^{nw} =0$.}
\begin{align}\label{GEquationM}
\tau_{(r} G^t_{mn)} &= \tau_s G^s_{(mn} \delta^t_{r)} - \frac{1}{2} \bar h_{(mn}  F_{r)s} h^{st} \, ,
\end{align}
with 
\begin{align}\label{defF}
F\equiv d \tau \, ,
\end{align}
characterising the spacetime torsion. The derivation of \eqref{RNC} and \eqref{GEquationM} from the geodesic equation of a particle evolving in a TNC background is shown in appendix \ref{appendixGeo}. We reproduce below the connection for a  generic TNC geometry \cite{connection1,connection2}
\begin{align}
\begin{split}\label{TNCconnection}
\Gamma^m_{rs} \equiv - \hat \upsilon^m \partial_r \tau_s + \frac{1}{2} h^{m t} \left(\partial_r \bar h_{s t} + \partial_s \bar h_{r t} - \partial_t \bar h_{r s} \right)\, .
\end{split}
\end{align}
It is compatible with the metrics $\tau_m$ and $h^{m n}$ and exhibits a torsion component $T^m_{\hphantom{m} r s} \equiv 2 \Gamma^m_{ [rs]}=- 2 \hat \upsilon^m \partial_{[r} \tau_{s]} = -\hat v^m F_{rs}$. While it is of course possible to proceed in the computation by using the connection $\Gamma^t_{mn}$, the solution to the geodesic equation \eqref{RNC} suggests  that a more natural connection to consider will be the one given by
\begin{align}\label{digamma1}
\mathring{\Gamma}^t_{mn} &\equiv \Gamma^t_{mn}+\frac{1}{2}\hat{\upsilon}^t  F_{mn}+G^t_{mn}.
\end{align}
This new connection is symmetric and $U(1)$ mass invariant. Although it is not compatible with $\tau_m$ and $h^{mn}$, the action of the new covariant derivative on these two tensors is quite simple:
\begin{align}\label{Dtauh}
\mathring{D}_m\tau_n&= \frac{1}{2}F_{mn}, \qquad\qquad\qquad \mathring{D}_rh^{mn}=a_{t}h^{t(m}\delta^{n)}_r.
\end{align}
Where $\mathring{D}$ denotes a covariant derivative with respect to the symmetric $U(1)$ mass invariant connection $\mathring{\Gamma}$,  the symbol $D$ will be reserved for the covariant derivative with respect to the standard TNC connection $\Gamma$.

From \eqref{quantumFields} and \eqref{RNC} it follows that 
\begin{align}
\begin{split}
\label{expansionX0}
\partial_\alpha  X^m &= \partial_\alpha X^m_0 + l_s \mathring{\nabla}_\alpha Y^m - l_s \left( \mathring{\Gamma}^m_{ts} \right)_0 Y^s \partial_\alpha X^t_0  - l^2_s \left( \mathring{\Gamma}^m_{rs}\right)_0 \mathring{\nabla}_\alpha Y^r Y^s \\ &\hphantom{=} - \frac{l^2_s}{2} \left[ \partial_t \mathring{\Gamma}^m_{rs} - 2 \mathring{\Gamma}^m_{ns} \mathring{\Gamma}^n_{tr} \right]_0 Y^r Y^s \partial_\alpha X^t_0 + \mathcal{O}\left( l^3_s \right)\, ,
\end{split}
\end{align}
where $\mathring{\nabla}_\alpha \equiv  \partial_\alpha X^n_0 \mathring{D}_n Y^m = \partial_\alpha Y^m + \left( \mathring{\Gamma}^m_{rs} \right)_0 \partial_\alpha X^r_0 Y^s $ is the pullback of the TNC spacetime covariant derivative $\mathring{D}_n$ onto the worldsheet. To compute $\bar S[\Psi_0,\Psi](0)$ we will also need the quantum expansion of the non-linear couplings $\bar h_{m n}(X), \bar B_{mn}(X), \aleph_m(X)$ and $\tau_m (X)$. This can be achieved by noting that any vector $V_m (X)$ and tensor $W_{mn}(X)$ can be expanded as
\begin{align}
\begin{split}\label{expansions}
W_{mn} &= \left(W_{mn}\right)_0 + \left( \partial_r W_{mn} \right)_0 l_s  Y^r + \frac{1}{2} \left( \partial_r \partial_s W_{mn} - \mathring{\Gamma}^t_{rs} \partial_t W_{mn}   \right)_0 l^2_s  Y^r Y^s \, ,\\
V_m &= \left( V_m \right)_0 + \left( \partial_r V_m \right)_0 l_s Y^r + \frac{1}{2} \left(  \partial_r \partial_s V_m - \mathring{\Gamma}^t_{rs} \partial_t V_m \right)_0 l^2_s Y^r Y^s \, ,
\end{split}
\end{align}
where we have made use of \eqref{RNC}. It is also straightforward to show that the pullback of any vector $V_m(X)$ and tensor $W_{mn}(X)$  can be written in the TNC covariant form 

\begin{align}
\begin{split}\label{pullbackExpansions1}
\frac{W_{\alpha \beta}}{l_s^2} &=  W_{mn} \mathring{\nabla}_\alpha Y^m \mathring{\nabla}_\beta Y^n + \mathring{D}_s W_{mn}  \mathring{\nabla}_\alpha Y^mY^s \partial_\beta X^n_0+ \mathring{D}_s W_{mn}  \mathring{\nabla}_\beta Y^n Y^s \partial_\alpha X^m_0  \\
&\hphantom{=} + \frac{1}{2}\left( \mathring{D}_r \mathring{D}_s W_{mn} + \mathring{R}^t_{srm} W_{tn}  + \mathring{R}^t_{srn} W_{mt} \right)Y^rY^s\partial_{\alpha}X_0^m\partial_{\beta}X_0^n+ \mathcal{O}\left(l_s\right) \, ,
\end{split}
\end{align}

\begin{align}
\begin{split}\label{pullbackExpansions2}
\frac{V_\alpha}{l_s^2} &= \frac{V_m \mathring{\nabla}_\alpha Y^m + \mathring{D}_r V_m Y^r \partial_\alpha X^m_0 }{l_s} \\
&\hphantom{=}+ \left[  \mathring{D}_m V_n Y^m \mathring{\nabla}_\alpha Y^n  + \frac{1}{2}  \left(\mathring{D}_r \mathring{D}_s V_m + \mathring{R}^t_{srm} V_t\right) Y^rY^s\partial_{\alpha}X_0^m  \right] + \mathcal{O}\left(l_s\right) \, ,
\end{split}
\end{align}
where $\mathring{R}^t_{srm} \equiv \partial_r \mathring{\Gamma}^t_{ms} - \partial_m \mathring{\Gamma}^t_{rs} + \mathring{\Gamma}^t_{rw} \mathring{\Gamma}^w_{ms} - \mathring{\Gamma}^t_{mw} \mathring{\Gamma}^w_{rs}$ is the Riemann tensor defined  in the usual way from the connection \eqref{digamma1} and where to avoid cluttering we have dropped the zero index on the background tensor fields. Making use of \eqref{pullbackExpansions1} and \eqref{pullbackExpansions2} we can rewrite the Polyakov action \eqref{Polyakov} in the TNC covariant way,  see apendix \ref{appendixRecombine} for its derivation, 

\begin{align}
\begin{split}\label{S0}
	\bar S_0 &= -  \int \frac{d^2\sigma e}{4\pi}\left[ \bar h_{mn} \mathring{\nabla}_\alpha Y^m \nabla^\alpha Y^n - \bar \Lambda_+ e^\beta_- \left(\mathring{\nabla}_\beta \hat H +  \mathring{\nabla}_\beta \left(\tau_m Y^m\right) \right)- \bar \Lambda_-e^\beta_+ \left(\mathring{\nabla}_\beta \hat H -  \mathring{\nabla}_\beta \left(\tau_m Y^m\right)\right) \right] \\
	&\hphantom{=} - \int \frac{d^2 \sigma e}{4 \pi} \left[ \bar \Lambda_+ Y^r \left(F_{mr} + \mathfrak{h}_{mr} \right) e^\beta_- \partial_\beta X^m_0 - \bar \Lambda_- Y^r \left( F_{mr} - \mathfrak{h}_{mr}  \right) e^\beta_+ \partial_\beta X^m_0 \right] \\
	&\hphantom{=} - \int \frac{d^2 \sigma e}{4 \pi}\left[ \left(\gamma^{\alpha \beta} A_{smn} + \epsilon^{\alpha \beta} \bar A_{smn} \right)  Y^s \mathring{\nabla}_\alpha Y^m \partial_\beta X^n_0 + \frac{1}{2} \left( \Delta \lambda^\beta F_{mn} - \Sigma \lambda^\beta \mathfrak{h}_{mn} \right) Y^m \mathring{\nabla}_\alpha Y^n  \right] \\
	&\hphantom{=} -\int \frac{d^2 \sigma e}{4 \pi} \left[ \left( \gamma^{\alpha \beta} C_{r s mn} + \epsilon^{\alpha \beta} \bar C_{rsmn} \right) Y^r Y^s \partial_\alpha X^m_0 \partial_\beta X^n_0 + \left( \Delta\lambda^\alpha B_{rsm} + \Sigma \lambda^\alpha \bar B_{rsm} \right) Y^r Y^s \partial_\alpha X^m_0 \right]\,  ,
\end{split}
\end{align}
where  $\hat H= \bar H + \aleph_m  Y^m$, $H=d\bar B, F= d\tau$, $\mathfrak{h}=d\aleph$, $\Delta \lambda^\beta \equiv \lambda^0_- e^\beta_+ - \lambda^0_{+} e^\beta_-$,  $\Sigma \lambda^\alpha \equiv  \lambda^0_- e^\beta_+ + \lambda^0_{+} e^\beta_-$ and where the coefficients $\{A,\bar A,B,\bar B,C,\bar C \}$   are given by

\begin{align}\label{curvedCoeff}
	\begin{split}
		A_{smn}&=2\mathring{D}_s \bar h_{mn}\, , \\
		\bar A_{smn} &=H_{smn}\, , \\
		C_{rsmn} &=   \frac{1}{2} \mathring{D}_r \mathring{D}_s \bar h_{mn} +  \mathring{R}^t_{\hphantom{t}(rs)(m} \bar h_{n)t} \, , \\
		\bar C_{rsmn} &= \frac{1}{2} \mathring{D}_r H_{smn} \, , \\
		B_{rsm} &=\frac{1}{2}\mathring{D}_r F_{sm} \, ,\\ 
		\bar B_{rsm} &=  - \frac{1}{2} \mathring{D}_r {\mathfrak{h}}_{sm} \, .
	\end{split}
\end{align}

We note that \eqref{S0} is manifestly invariant under the $U(1)_\mathcal{B}$ zero and one form transformations as it is written exclusively in terms of $\mathfrak{h}$ and $H$ instead of $\aleph$ and $\bar B$. Ideally one would like to do the same for the $U(1)$ mass symmetry, i.e. express the action in terms of the field strength of $m$, however this would give us an action which is manifestly $U(1)$ mass invariant, but not manifestly Galileian invariant. Although the action will be kept in its explicit Galilean invariant form, as written in \eqref{S0}, it can be shown that it is still invariant under the $U(1)$ mass symmetry after making use of the classical equations of motion \eqref{boxfin} as well as the transformation rules for the quantum Lagrange multipliers, derived from \eqref{lambdaU1}:
 \begin{align} \label{LambdaU1}
\delta \Lambda_{\pm} = \mp e^\alpha_{\pm}\left(\mathring{D}_r \sigma\mathring{\nabla}_\alpha Y^r+\mathring{D}_r\mathring{D}_s\sigma Y^r\partial_{\alpha}X_0^s\right)\, .
\end{align}
The preservation of this symmetry at the quantum level is then expected to be non-trivial.
\subsection{Weyl invariance at one loop}\label{oneLoop}
\label{betaFunctions}

From \eqref{S0} we observe that $\Gamma[\Psi_0](0)$ is a free theory with a background dependent  normalization for the kinetic and mass terms. Nevertheless, since we are looking at contributions up to $\mathcal{O}\left(D^2\right)$ in target spacetime derivatives we can treat \eqref{oneLoopAction} perturbatively as long as we can renormalize the $\mathcal{O}\left(D^0\right)$ appropriately. One can move these background dependent norms to terms higher order in spacetime derivatives through the following coordinate transformation
\begin{align}
\begin{split}\label{flatDecomposition}
 Y^m &= - \bar  \upsilon^m \left(\tau_s Y^s  \right) + e^m_i \left(\delta^{ij} e^r_j \bar h_{r s} Y^s \right)\equiv - \hat \upsilon^m \frac{Y^0}{\sqrt{2 \Phi}} + e^m_i Y^i \equiv e^m_I Y^I, \\
 \hat H &= \frac{H}{\sqrt{2 \Phi}}, \quad \quad \bar \Lambda_\pm  = \sqrt{2 \Phi} \Lambda_{\pm}\, ,
\end{split}
\end{align}
%
with $\Phi$ defined in (\ref{Phi}), $Y^I=\{Y^0,Y^i \}$ and the normalizations are judiciously chosen such that the normalization of the first term in (\ref{S0}) becomes canonical, i.e. it yields the first two terms in the zeroth order action below. To see this one needs to use the identity $\bar h_{mn} \hat \upsilon^n = 2\Phi \tau_m$ and in particular we can identify the spacetime inverse vielbeins $e^m_I=\{\frac{-\hat \upsilon^m}{\sqrt{2 \Phi}}, e^m_i \}$ satisfying $\bar h_{mn} e^m_I e^n_J = \eta_{IJ}$. The effective action $S_0$ is now expressed in terms of flat indices and can be expanded as
\begin{align}
\begin{split} \label{flatSplitting}
S_0 &= S^{[0]}_0 + S^{[1]}_0 + S^{[2]}_0 \, , 
\end{split}
\end{align}
with $S^{[a]}_0$ denoting the $\mathcal{O}\left( D^a \right)$ in target spacetime derivatives. In particular the $\mathcal{O}\left(D^0 \right)$ action is given by the free action with constraints 
\begin{align}
	\begin{split}\label{actionS0}
		S^{[0]}_0 &= - \int \frac{d^2 \sigma e }{4 \pi}  \left[\gamma^{\alpha \beta}\eta_{IJ} \partial_\alpha Y^I \partial_\beta Y^J  -   \Lambda_+ e^\beta_-\left( \partial_\beta H + \partial_\beta Y^0 \right) - \Lambda_- e^\beta_+ \left(\partial_\beta H  - \partial_\beta Y^0 \right)  \right]\, . \\ 
	\end{split}
\end{align}
Assuming a diffeomorphism invariant measure the path integration over the fields $\{ Y^m, \bar \Lambda, \bar H \}$ can be changed to an integration over $\{Y^0, Y^i, \Lambda, H \}$. After this change of coordinates, the following propagators for $S^{[0]}_0$ can be constructed 

\begin{align}
	\begin{split}\label{propagators}
		\langle Y^I(\sigma) Y^J(\sigma') \rangle_0 &= \Delta_2 \left(\eta^{IJ} + \delta^I_0 \delta^J_0 \right) \ln \left( |\Delta \sigma|^2 \right) \, , \\
		\langle Y^I(\sigma) \Lambda_{\pm}(\sigma') \rangle_0 &= \delta^I_0 \frac{\mp 2 \Delta_2}{ \left(\sigma - \sigma'  \right)_\pm} \, ,\\
		\langle H(\sigma) \Lambda_{\pm} (\sigma') \rangle_0 &= \frac{-2 \Delta_2}{\left(\sigma - \sigma' \right)_{\pm}} \, ,\\
		\langle \Lambda_{\pm} (\sigma) \Lambda_{\pm}(\sigma') \rangle_0 &= \frac{4 \Delta_2}{(\sigma -\sigma')_\pm}\, , \\
		\langle \Lambda_+(\sigma) \Lambda_- (\sigma') \rangle_0 &= - 4 \pi \Delta_2 \delta(\sigma - \sigma')\, ,
	\end{split}
\end{align}
where $\langle \rangle_0$ denotes the correlation function computed with respect to the action $S^{[0]}_0$ and where $\Delta_2$ is an unimportant overall factor. At first and second order in covariant derivatives we can perform the further decomposition 

\begin{align}
	\begin{split}\label{TildeDecomp}
		S^{[1]}_0 &= \mathcal{S}_1 + \mathcal{\tilde S}_1 \, ,\\ 
		S^{[2]}_0 &= \mathcal{S}_2 + \mathcal{\tilde S}_2 \, ,
	\end{split}
\end{align}
where we make a distinction between the contributions coming directly from coefficients $\{A,\bar A,C,\bar C, B, \bar B \}$ and the contributions coming from the non-compatibility of the vielbeins $\{\frac{-\hat \upsilon^m}{\sqrt{2 \Phi}},e^m_i \}$ by considering the former in $\mathcal{S}$ and the latter in $\mathcal{\tilde S}$. In detail we find for the $\mathcal{S}$ components 

\begin{align}
			\mathcal{S}_1 &=  - \int \frac{d^2 \sigma e}{4 \pi} \left[ \bar \Lambda_+ Y^I e^r_I \left(F_{mr} + \mathfrak{h}_{mr} \right) e^\beta_- \partial_\beta X^m_0 - \bar \Lambda_- Y^I e^r_I \left( F_{mr} - \mathfrak{h}_{mr}  \right) e^\beta_+ \partial_\beta X^m_0 \right] \\
	\nonumber &\hphantom{=} - \int \frac{d^2 \sigma e}{4 \pi}\left[ \left(\gamma^{\alpha \beta} A_{smn} + \epsilon^{\alpha \beta} \bar A_{smn} \right) e^s_I e^m_J  Y^I \partial_\alpha Y^J \partial_\beta X^n_0 \right] \\\nonumber &\hphantom{=} - \int \frac{d^2 \sigma e}{4 \pi} \left[ \frac{1}{2} \left( \Delta \lambda^\beta F_{mn} - \Sigma \lambda^\beta \mathfrak{h}_{mn} \right)e^m_I e^n_J Y^I \partial_\alpha Y^J  \right] \, ,	\\
	\mathcal{S}_2 &=  -\int \frac{d^2 \sigma e}{4 \pi} \left[ \left( \gamma^{\alpha \beta} C_{r s mn} + \epsilon^{\alpha \beta} \bar C_{rsmn} \right) e^r_I e^s_J Y^I Y^J \partial_\alpha X^m_0 \partial_\beta X^n_0 \right] \\
			\nonumber &\hphantom{=} -\int \frac{d^2 \sigma e}{4 \pi}\left[ \left( \Delta\lambda^\alpha B_{rsm} + \Sigma \lambda^\alpha \bar B_{rsm} \right) e^r_I e^s_J Y^I Y^J \partial_\alpha X^m_0 \right]\,  ,
\end{align}
and for the $\mathcal{\tilde S}$ components 
\begin{align}
		\tilde S_1 &= -\int \frac{d^2 \sigma e}{4 \pi} \left[ \frac{\Lambda_+ e^\beta_- H_+ + \Lambda_- e^\beta_+ H_-}{2} \partial_\alpha X^m_0 \mathring{D}_m \ln \Phi + \left(2 \bar h_{rs} e^r_I \mathring{D}_m e^s_J \right) Y^I \partial_\alpha Y^J   \partial^\alpha X^m_0 \right] \, ,  \\
		\tilde S_2 &= - \int \frac{d^2 \sigma e}{4 \pi} \left[ \left(\bar h_{rs} \mathring{D}_m e^r_I \mathring{D} e^s_J \right) \partial_\alpha X^m_0 \partial^\alpha X^n_0 \right] \\ \nonumber 
		&\hphantom{=} - \int \frac{d^2 \sigma e}{4 \pi} \left[ \left(A_{s r n} \gamma^{\alpha \beta} + \bar A_{s r n} \epsilon^{\alpha \beta}\right) e^s_I \mathring{D}_m e^r_J  Y^I Y^J \partial_\alpha X^m_0 \partial_\beta X^n_0\right]  \\ \nonumber
		&\hphantom{=} - \int \frac{d^2 \sigma e}{4 \pi} \left[ \left(\frac{\Delta \lambda^\alpha F_{s r}}{2} - \frac{\Sigma \lambda^\alpha \mathfrak{h}_{sr}}{2} \right)e^s_I \mathring{D}_m e^r_J  Y^I Y^J \partial_\alpha X^m_0 \right] \, ,
\end{align}
where we have explicitly broken the covariance by using $\mathring{\nabla}_\alpha Y^I = \partial_\alpha Y^I + \omega^{I}_{\hphantom{i}J \alpha} Y^J$ with $\omega^{IJ}_{\hphantom{IJ}\alpha}$ the spin connection\footnote{The spin connection is not  gauge invariant and consequently it will not contribute to the beta functions.}, and where the covariant derivative $\mathring{D}_me^t_I$ is taken only with respect to the curved spacetime indices. The effective action  \eqref{oneLoopAction} can be now be treated perturbatively and its corresponding Weyl variation, \eqref{variationEffective},  can be computed as 
\begin{align}
\begin{split}\label{variationExplicit}
\delta_\psi \bar \Gamma[\Psi_0](0) &= \delta_\psi \left \langle S^{[1]} + S^{[2]} \right \rangle_0 + \frac{i}{2} \delta_\psi \left \langle S^{[1]} S^{[1]} \right \rangle_0 \\ &\hphantom{=} - i \delta_\psi \left \langle S^{[1]} \right \rangle_0 \left \langle S^{[1]} \right \rangle_0  - i \delta_\psi \log(Z_0 Z_{FP})  + \mathcal{O}\left(D^3\right),
\end{split}
\end{align}
where  $Z_{FP}$ is the partition function for the Fadeev-Popov ghosts arising from the gauge fixing procedure, see appendix \ref{ghosts}, and where $Z_0$ denotes the partition function with respect to the action $S^{[0]}_0$. By dimensional considerations we expect $\delta_\psi  \log(Z_0  Z_{FP}) = c_T \mathcal{R}$ with $c_T$ a proportionality constant\footnote{At one loop level this is the only contribution to the anomaly proportional to $\mathcal{R}$.}. The coefficient $c_T$ is independent of the background fields and depends only on the dimensionality of the TNC spacetime. Therefore, as in the case of the ordinary string, the requirement $c_T=0$  fixes the dimensionality of the background geometry. This is the requirement of invariance under conformal reparametrizations, hence the quantum consistency of the theory in the absence of extra dynamical fields. We find that the requirement $c_T=0$ fixes the critical dimension of the $d+1$ dimensional TNC geometry to be 
\begin{equation}
\label{critical}
d_c+1=25\, .
\end{equation}
 The details of this calculation are presented in Appendix  \ref{criticalDimension}. This result is somewhat expected, as quantum consistency of the ordinary bosonic string sets $d+2=26$ and we obtain the TNC geometry by reduction of this 26 dimensional background on a null direction. Nevertheless, it is still a non-trivial result, as we cannot find a simple argument as to why quantization and null reduction should commute. Taking the dimension to be critical, we expect the right hand side of \eqref{variationExplicit} to take the form 
\vspace{2mm}
\begin{align}
\begin{split}\label{defBeta}
\delta_\psi \bar \Gamma[\Psi_0](0) &= -  \int d^2 \sigma \frac{\psi}{4 \pi} \left[ \beta_{r s} \eta^{\alpha \beta} \partial_\alpha X^r \partial_\beta X^s + \bar \beta_{rs} \epsilon^{\alpha \beta}\partial_\alpha X^r \partial_\beta X^s  \right. \\ & \hphantom{=} \hphantom{-\int d^2 \sigma \frac{\psi}{2\pi}[[[[ } \left. + \beta_m \Delta \lambda^\beta \partial_\beta X^m_0    + \bar \beta_m  \Sigma \lambda^\beta \partial_\beta X^m_0 +\beta \lambda^0_+ \lambda^0_-\right]\, .\end{split}
\end{align}
where  $\{\beta,\beta_{rs}, \bar \beta_{rs},\beta_m,\bar \beta_{m} \}$ will correspond to the beta functions. We will exemplify the computation of the beta functions by taking the background solution to be $\partial_\alpha X^m_0=0$ so that we can easily compute the scalar beta function $\beta$. Under this assumption and making use of \eqref{variationExplicit}, \eqref{actionS0}, and \eqref{TildeDecomp} we find 

\begin{align}
	\begin{split}\label{var1Main}
		-\delta_\psi \bar \Gamma \left[ \Psi_0 \right] (0) &=  \delta_\psi \int \frac{d^2 \sigma e}{4 \pi} \left[  \frac{1}{2} \left(\Delta \lambda^\beta F_{mn} - \Sigma \lambda^\beta \mathfrak{h}_{mn} \right) \right]\Delta^{mn} +\\
		&\hphantom{=}  \delta_\psi \int \frac{d^2\sigma d^2\sigma'i e^2}{64 \pi^2}  \left[ \left(F_{rs} F_{tw} - \mathfrak{h}_{rs} \mathfrak{h}_{tw} \right) \gamma^{\alpha \beta} +F_{rs}\mathfrak{h}_{tw}\epsilon^{\alpha \beta} \right] \lambda_+ \lambda_- \gamma^{\alpha \beta}\Delta^{rstw}_{\alpha \beta} \, ,
		\end{split}
\end{align} 
where for simplicity we have defined 

\begin{align}\label{propMain}
	\Delta^{m n}_\alpha(\sigma) &\equiv e^m_I e^n_J \langle Y^I(\sigma) \partial_\alpha Y^J(\sigma) \rangle_0 \, , \\
	\Delta^{rstw}_{\alpha \beta}(\sigma, \sigma') &\equiv e^r_Ie^s_J e^t_K e^w_L \langle Y^I(\sigma) \partial_\alpha Y^J(\sigma) Y^K (\sigma') \partial_\beta Y^L(\sigma') \rangle_0 \, .
\end{align}

The propagators in \eqref{propMain} can be computed by making use of the zeroth order action \eqref{actionS0}, and in particular the following identities follow from it

\begin{align}
	\begin{split}\label{identitiesMain}
		\delta_\psi \int d^2 \sigma J^{\alpha}_{r s } \Delta^{r s}_\alpha(\sigma) &= - \frac{1}{2} \int d^2 \sigma e \psi \partial_\alpha \left( h^{rs} J^\alpha_{rs} \right) \, , \\
		\delta_\psi \int d^2 \sigma d^2 \sigma' J^{\alpha \beta}_{rstw} \Delta^{rstw}_{\alpha \beta}(\sigma, \sigma') &= (-2 \pi i) \int d^2 \sigma e \psi J^{\alpha \beta}_{rstw}h^{rt}h^{sw} \gamma_{\alpha \beta} \, ,
	\end{split}
\end{align}
where $\{J^\alpha_{rs},J^{\alpha \beta}_{rstw} \}$ are arbitrary tensors. By using \eqref{identitiesMain} and \eqref{var1Main} we finally find $\beta$ to be

\begin{align}\label{betaScalar}
	\beta &= \frac{1}{4} \left(\mathfrak{h}_{rs}\mathfrak{h}_{tw} - F_{rs} F_{tw} \right) h^{rt} h^{sw} \, .
\end{align}

Our analysis depends on the existence of a solution $G^t_{rs}$ to the geodesic equation \eqref{GEquationM}. It is easy to show that such solution exists as long as the torsion is twistless, namely that it satisfies the constraint
\begin{align}\label{twistless}
	F_{r s} h^{rt} h^{sw} &= 0\, ,
\end{align}
with corresponding solution to the geodesic equation given by

\begin{align}\label{gSolution}
	G^t_{mn} &= \frac{1}{2}\bar h_{mn} F_{rs} \hat \upsilon^r h^{st} \, .
\end{align}

For the rest of this work we will use \eqref{gSolution} and assume \eqref{twistless} holds for the TNC background. This requirement together with the Weyl invariance condition $\beta=0$  implies that, just as $F$, the field strength $\mathfrak{h}$ is forced to be twistless. This condition can be made explicit by expressing $F$ and $\mathfrak{h}$ in terms of the decomposition \cite{Dieter}
 
 \begin{align}
 	\begin{split}\label{twistlessDecomp}
 		F_{rs}&\equiv a_r \tau_s - \tau_r a_s \, , \\
 		\mathfrak{h}_{rs}&\equiv \mathfrak{e}_r \tau_s - \tau_r \mathfrak{e}_s \, ,
 	\end{split}
 \end{align} 
with $a_r= \hat \upsilon^t F_{tr}$ the acceleration and $\mathfrak{e}_r = \hat \upsilon^t \mathfrak{h}_{tr}$ an electric field, and where both vectors satisfy $a_t \hat \upsilon^t =\mathfrak{e}_t \hat \upsilon^t=0$. For simplicity and from now on we will assume \eqref{twistlessDecomp} holds for the computation of the remaining beta functions. It is important to note we should think of \eqref{twistless} not as an equation of motion arising from Weyl invariance but rather as a constraint to ensure both general covariance and $U(1)$ mass invariance at the quantum level.

Taking $\partial_\alpha X^m_0$ satisfying \eqref{boxfin} and following a similar procedure as the one just outlined for the computation $\beta$, the remaining beta functions are found to be

\begin{align}\label{beta1M}
 \beta_m &= \left[\frac{1}{2} \mathring{D}\cdot a + \left( \frac{d_c}{4} + \frac{1}{2} \right) a^2 - \mathfrak{e}^2 - a \cdot \mathring{D}\phi \right] \tau_m \, ,\\
 \label{beta2M}\bar \beta_m &= - \left[ \frac{1}{2} \mathring{D}\cdot \mathfrak{e} + \frac{d_c}{4} a \cdot \mathfrak{e} - \mathfrak{e} \cdot \mathring{D}\phi  \right] \tau_m \, , \\
 \label{beta3M}\beta_{mn} &= - \mathring{R}_{mn} + \frac{1}{4}H_{rs(m} H_{n)tw} h^{rt} h^{sw}- 2 \mathring{D}_{(m} \mathring{D}_{n)} \phi - \mathfrak{e}_r h^{rs} \left( \Delta_T \right)^t_{(m} H_{n)ts}  \\ \nonumber
 &\hphantom{=}    + \frac{\mathfrak{e}^2 \left(2 \Phi \tau_m \tau_n + \bar h_{mn} \right) - \mathfrak{e}_m \mathfrak{e}_n}{2} -\beta_t \hat \upsilon^t \bar h_{mn} \, , \\
 \label{beta4M}\bar \beta_{mn} &= \frac{1}{2} h^{rs} \mathring{D}_r H_{smn} + \frac{d_c}{4} a_r h^{rs}H_{smn}  - \left(\Delta_S \right)^t_{[m} \mathring{D}_{n]} \mathfrak{e}_t+\left( \Delta_T\right)^r_{[m} \mathring{D}_{n]} \mathfrak{e}_r  +a_{[m} \mathfrak{e}_{n]} 
 \\ \nonumber &\hphantom{=}  + \frac{\mathring{D}_t \upsilon^t}{2} \mathfrak{h}_{mn} - \left(\hat \upsilon^r \mathfrak{h}_{mn} + h^{rp} H_{p m n} \right) \mathring{D}_r \phi\, ,
 \end{align}
with $a^2=a_r a_s h^{rs}$, $\mathfrak{e}^2=\mathfrak{e}_r \mathfrak{e}_s h^{rs}$, $\mathring{R}_{mn}$ the Ricci tensor, $\cdot$ denoting an inner product with respect to $h^{rs}$, and where the time projector $\left(\Delta_T\right)^t_m$ and the space projector $\left( \Delta_S \right)^t_m$ are defined as

\begin{align}
	\begin{split}
		\left( \Delta_T \right)^t_m &= - \hat \upsilon^t \tau_m \, , \qquad \qquad 		\left( \Delta_S \right)^t_m = h^{tp}\bar h_{p m}\, ,
	\end{split}
\end{align}
satisfying the projector identities 

\begin{align}
	\begin{split}
		\left( \Delta_T \right)^t_m + \left( \Delta_S \right)^t_m &= \delta^t_m \\
		\left( \Delta_{T/S} \right)^t_m \left( \Delta_{T/S} \right)^m_w &= \left( \Delta_{T/S} \right)^t_w\\ \left( \Delta_T \right)^t_m \left( \Delta_S \right)^m_w &= 0
			\end{split}
\end{align}
The details of the derivation of \eqref{beta1M}-\eqref{beta4M}  can be found in appendix \ref{betaF}. The Weyl invariance of the theory at one loop will follow from the vanishing of the beta functions. These constraints will be interpreted as the gravitational equations of motion for the TNC background, such equations are discussed in the following section. Before finalizing this section we comment on the $U(1)$ mass covariance of the beta functions \eqref{beta1M}-\eqref{beta4M} by noting that 

\begin{align}
	\begin{split}\label{betaVar}
	 \delta_\sigma \beta &=0 \, ,\\
		\delta_{\sigma} \beta_m &= 0 \, ,\\
		\delta_\sigma \bar \beta_m &= 0\, , \\
		\delta_\sigma \beta_{mn} &=  2 \left(\beta_t \hat \upsilon^t \right) \tau_{(m}\mathring{D}_{n)} \sigma \, , \\
		\delta_\sigma \bar \beta_{mn} &=  2 \left(\bar \beta_t \hat \upsilon^t \right)  \tau_{[m}\mathring{D}_{n]} \sigma \, ,
	\end{split}
\end{align}
where the transformation rules \eqref{hbTransf}, \eqref{trbarups}, \eqref{trPhi}, and \eqref{HTransf} have been used. From \eqref{betaVar} we can note that the vanishing of the beta functions is a $U(1)$ mass invariant condition.
\subsection{TNC equations of motion}
\label{TNCequations}
 
The gravitational equations for the TNC background will arise from the condition \eqref{twistless}, and by setting \eqref{betaScalar},\eqref{beta1M}-\eqref{beta4M} to zero. The resulting equations can be  categorized into two twistless constraints:

\begin{align}
	   \label{beta1} F_{rs}&= a_r \tau_s - \tau_r a_s \, , \\
 		\label{beta2} \mathfrak{h}_{rs}&= \mathfrak{e}_r \tau_s - \tau_r \mathfrak{e}_s \, ,
\end{align}
two scalar equations: 

\begin{align}
	\label{beta3}  D\cdot a + a^2 &= 2 \mathfrak{e}^2 + 2 \left(a \cdot D\phi \right) \, ,\\
 \label{beta4} D\cdot \mathfrak{e} &= 2 \left(\mathfrak{e} \cdot D\phi\right)  \, , 
\end{align}
and two tensor equations: 		
				\begin{align}
  \label{beta5}  R_{(mn)} - \frac{H_{rs(m} H_{n)tw} h^{rt} h^{sw}}{4} + 2 D_{(m} D_{n)} \phi &= \frac{\mathfrak{e}^2 \left(2 \Phi \tau_m \tau_n - \bar h_{mn} \right) - \mathfrak{e}_m \mathfrak{e}_n}{2}   - \mathfrak{e}_r h^{rs} \left( \Delta_T \right)^t_{(m} H_{n)ts}   \nonumber \\
  &\hphantom{=} +\left(\Delta_S \right)^t_{(m} D_{n)}a_{t} + \frac{a_m a_n}{2}    -a^2\, \Phi \tau_m\tau_n  \, , \\
 \label{beta6} \frac{1}{2} h^{rs} D_r H_{smn} - h^{rp}H_{pmn}D_r\phi &= \left(\Delta_S \right)^t_{[m} D_{n]} \mathfrak{e}_t -\left( \Delta_T\right)^r_{[m} D_{n]} \mathfrak{e}_r  -a_{[m} \mathfrak{e}_{n]}
   \nonumber  \\  &\hphantom{=}  - \frac{1}{2} a_r h^{rs}H_{smn}  + \left( \hat \upsilon^t D_t \phi - \frac{D_t \upsilon^t}{2} \right)\mathfrak{h}_{mn} \, .
 \end{align}
 
In \eqref{beta1}-\eqref{beta6} we have used the original TNC connection \eqref{TNCconnection} and used $D$ to denote its corresponding covariant derivative. The Ricci tensor associated to the standard TNC connection can be read off from the following relation
\begin{align}
\begin{split}
\mathring{R}_{mn}-R_{mn} &= -\frac{1}{2}a_m a_n -D_n a_m + \frac{1}{2}\left(D\cdot a + a^2\right)\bar{h}_{mn} +\left(\Delta_T\right)^r_{[m}D_r a_{n]} +\left(\Delta_T\right)^r_{(m}D_{n)} a_{r} \\
&\hphantom{=}\ \, -\frac{1}{2}D_r \hat{\upsilon}^r F_{mn}+a^2\Phi \tau_m\tau_n.
\end{split} 
\end{align}
 Notice that it is not symmetric in the presence of torsion and, as discussed earlier, the TNC connection is not $U(1)$ mass invariant when torsion is non-vanishing. Consequently the $U(1)$ mass invariance of equations \eqref{beta1}-\eqref{beta6} is harder to verify in this form, however from \eqref{betaVar} we know they are indeed invariant. We can also note that unlike the expressions \eqref{beta1M}-\eqref{beta4M}, where the $U(1)$ mass invariant connection has been used, equations \eqref{beta1}-\eqref{beta6} have no explicit dependence on the critical dimension $d_c$. At this point it is convenient to introduce the extrinsic curvature tensor $\mathcal{K}_{mn}$ as \cite{ActionPrinciple}

\begin{align}
	\mathcal{K}_{mn} \equiv - \frac{1}{2} \mathcal{L}_{\hat \upsilon} \bar h_{mn} = -\frac{1}{2} \left[\hat \upsilon^t D_t \bar h_{mn} + \bar h_{mt} D_n \hat \upsilon^t + \bar h_{nt} D_m \hat \upsilon^t -4 \Phi a_{(m}\tau_{n)} \right] \, ,
\end{align}
and use the TNC identity

\begin{align}
	D_m \bar h_{rs} &= 2 \tau_{(r} \bar h_{s)p} D_m \hat \upsilon^p - 2 \tau_r \tau_s D_m \Phi \, ,
\end{align}

\noindent
to derive the following contractions of the extrinsic curvature  
 
\begin{align}\label{KCont}
\begin{split}
 h^{rs} \mathcal{K}_{rs} &= - D_t  \hat \upsilon^t \, ,\\
	 \mathcal{K}_{rs}\mathcal{K}_{tw} h^{rt} h^{sw} &= D_m \hat \upsilon^n D_n \hat \upsilon^m  \, , 
\end{split}
\end{align}
We can then see that $\mathcal{K}_{rs}h^{rs}$ shows up in the antisymmetric beta function \eqref{beta6}. To further show the role of $\mathcal{K}_{mn}$ in equations \eqref{beta1}-\eqref{beta6} it is instructive to look at the time-time projection of equation \eqref{beta5} to write down Newton's law in a general TNC spacetime. For this it will be necessary to use the $\hat \upsilon^m \hat \upsilon^n$ projection of the TNC Bianchi identity  \eqref{Bianchi}

\begin{align}
	\begin{split}
			\hat \upsilon^n D_n \hat \upsilon^t &= h^{ts} \left(D_s \Phi + 2 a_s \Phi \right) \, ,\\
	\end{split}
\end{align}
the scalar equation \eqref{beta3}, and the extrinsic curvature contractions  \eqref{KCont} to find that Newton's law takes the form
\begin{align}
	\begin{split}\label{NewtonLaw}
		D^2 \Phi  + 3 \left(a \cdot D\Phi\right) +  m^2_\Phi\,\Phi  &=  \rho_{\mathcal{\kappa}}+ \rho_{\text{m}} \, ,
	\end{split}
\end{align}  
with  $D^2 \equiv h^{rs}D_r D_s$, and where the Newton's potential mass $m_\Phi^2$, matter density $\rho_{\text{m}}$, and curvature density $\rho_{\mathcal{K}}$ are defined as 

\begin{align}
	m^2_{\Phi}&\equiv a^2+2\,\mathfrak{e}^2 + 4\,a \cdot D\phi  \, , \\
	\rho_{\mathcal{K}} &\equiv \mathcal{K}_{rs} \mathcal{K}_{tw}h^{rt}h^{sw} - \hat \upsilon^n D_n \left( \mathcal{K}_{rs} h^{rs}\right) \, , \\
	\rho_{\text{m}} &\equiv \frac{1}{4} \hat \upsilon^m \hat \upsilon^n h^{rt} h^{sw} H_{rsm} H_{twn} - 2 \hat \upsilon^m \hat \upsilon^n D_m D_n \phi \, .
\end{align}

From \eqref{NewtonLaw} we can observe that the extrinsic curvature enters Newton's law in the form of a matter density distribution. In contrast we can note that the presence of torsion modifies considerably the classical gravitational equation of motion by adding both a mass term\footnote{From \eqref{beta3} we note that whenever torsion vanishes the electric field $\mathfrak{e}$ also vanishes and cosequently $m^2_\Phi=0$.} via its coupling with matter through \eqref{beta3}, and an advection term via the coupling $a \cdot D\Phi$. Equation \eqref{NewtonLaw} is nothing but the temporal trace of the Ricci tensor, however it is also instructive to compute its spatial trace $\mathcal{R}_S \equiv R_{mn} h^{mn}$ to find that

\begin{align}
	\begin{split}\label{spatialRicci}
		\mathcal{R}_S &= \frac{1}{4} H^2_S -2 D^2 \phi + \frac{m^2_\Phi}{2} - \frac{\left(d_c -1 \right) \mathfrak{e}^2}{2} -a^2\, ,
	\end{split}
\end{align}
with $H^2_S \equiv H_{rsm} H_{twn} h^{rt} h^{sw} h^{mn}$. In addition  the electric Maxwell equation \eqref{beta4} reduces to Gauss' law only for a vanishing dilaton while the two-form Maxwell equation \eqref{beta6} is not only sourced by $\phi$ and $\mathfrak{e}$ but also by torsion via the coupling $a_r h^{rs} H_{smn}$. 

Finally we mention a few properties of torsion and what role it plays in the equations of motion. First of all we recall that the conditions $\Gamma^t_{[mn]}=0$ and $a_m=0$ are completely equivalent as long as torsion is forced to be twistless. In the torsionless case (i.e. when the acceleration vanishes) we notice that the electric field $\mathfrak{e}$ is also forced to vanish. On the other hand, a non vanishing electric field forces torsion  and the Kalb-Ramond field strength to be non-zero. The first property can be read off explicitly from \eqref{beta3}, while the second one is a consequence of the $U(1)$ mass transformation \eqref{HTransf}. Hence in the absence of torsion the mass  and advection terms in Newton's law vanish,  yelding the more familiar Poisson equation
\begin{align}
	\begin{split}
		D^2 \Phi    &=  \rho_{\mathcal{\kappa}}+ \rho_{\text{m}} \, .
	\end{split}
\end{align}  
Lastly we notice that for vanishing torsion the TNC equations of motion assume the same form as the usual equations derived from bosonic string theory.

\section{Discussion and Outlook} \label{Conclusions}

We studied the non-linear sigma model for a bosonic string moving in torsional Newton-Cartan geometry at one-loop. Demanding Weyl invariance at this level yields the critical dimension of space-time and the equations of motion of the TNC background. We found $d_c=25$ for the critical dimension. The equations of motion for non vanishing torsion, three form field strength $H$, electric field $\mathfrak{e}$, and dilaton $\phi$ are given in \eqref{beta1}-\eqref{beta6}.

Our result for the critical dimension is not surprising as the classical TNC geometry is obtained by reduction of an ordinary Riemannian background on a null direction, and quantum Weyl invariance of a bosonic string on a Riemannian background requires $d=26$. However it is still non-trivial, as there is, a priori, no guarantee that the argument of null reduction carries over to the quantum regime. As seen from the calculation in Appendix \ref{criticalDimension}, the number 25 arises from quite a non-trivial calculation that involves the TNC ghost sector and the constraint equations. Our result, therefore is somewhat non-trivial and implies that null reduction and quantization are commuting operations.

To compute the beta functions of the theory we used the TNC geodesic equation to derive a system of normal coordinates $Y^m$ such that covariant results could be obtained. For this local coordinates to exist we needed the twistlessness constraint $\tau \wedge d \tau=0$ which also guarantees causality in this non-relativistic space-time \cite{Bergshoeff:2017dqq}. Moreover we introduced a new connection with the useful properties of being both symmetric and $U(1)$ mass invariant. The latter property was especially useful when checking the invariance of the beta functions. In fact, the $U(1)$ mass symmetry was used as a guiding principle to find the correct set of equations, as the beta functions would otherwise be dependent on some arbitrary coefficients, see \eqref{4pC}. Intuitively these coefficients are related to a choice of renormalization scheme and are uniquely fixed by requiring the $U(1)$ mass symmetry to be conserved at the quantum level.

The resulting equations of motion \eqref{beta1}-\eqref{beta6} for the TNC target spacetime are invariant under all TNC transformations as well as the corresponding 1-form and 2-form symmetries of the Kalb-Ramond fields. In the absence of torsion they take the form of the usual bosonic string equations of motion and yield the expected Newton's law for the gravitational potential $\Phi$. Once torsion is turned on Newton's law is modified accordingly with a mass term and an advection term for the gravitational potential $\Phi$ being generated.  

Our work can be improved and generalized in a number of ways. First, it is desirable to obtain the  ${\cal O}(l_s^2)$ contributions to the dilaton beta function. As mentioned above, this requires two-loop calculations on the worldsheet which can be done in the case of the bosonic string with relative ease but in our case there exist more than 20 contributions with different structures and  this computation becomes  a formidable task. Yet, it is a straightforward task and should be done in the near future. 
It is curious to compare our equations with the ones obtained from other effective approaches, such as the action principle proposed in \cite{ActionPrinciple}, and the large $c$ expansion of general relativity equations in \cite{Dieter}. Finally, it is very interesting  to ask  whether one can obtain Weyl invariant {\em sub-critical} TNC backgrounds with dimensionality less than 25 by searching for analogs of the linear-dilaton type geometry in the ordinary bosonic string case. In that case the slope of the linear dilaton cancels the ${\cal O}(l_s^0)$ contribution to the dilaton beta function hence lifting the condition $d=26$ and allowing for non-Lorentz invariant backgrounds with an arbitrary $2<d<26$. Since we already gave up  Lorentz invariance in the target spacetime, it is natural to ask if one can obtain subcritical TNC geometries with an analogous mechanism. To see if this is possible one will need the ${\cal O}(l_s^2)$ contributions to the dilaton beta function.

\begin{acknowledgements}

We thank Eric Bergshoeff, Jelle Hartong, Niels Obers, Ceyda Simsek, Dieter Van den Bleeken, Ziqi Yan, Troels Harmark, Gerben Oling, and Jan Rosseel for very useful discussions. This work is partially supported by the Netherlands Organisation for Scientific Research (NWO)
under the VIDI grant 680-47-518 and the Delta-Institute for Theoretical Physics (D-ITP), both
funded by the Dutch Ministry of Education, Culture and Science (OCW). DG is supported in part by CONACyT through the program Fomento, Desarrollo y Vinculacion de Recursos Humanos de Alto Nivel.
\end{acknowledgements}

\newpage 
\appendix

\section{Geodesic equation and normal coordinates in TNC geometry}\label{appendixGeo}

The action of a particle moving in a TNC background is given by \cite{Andringa:2012uz}

\begin{align}
\begin{split}
\mathcal{S}_{\text{part}} = \int d \lambda \frac{m}{2} \frac{\bar h_{m n} \dot x^m \dot x^n}{\tau_s \dot x^s} \, .
\end{split}
\end{align}
The geodesic equation can be obtained by minimising such action, the corresponding equations of motion are found to be 

\begin{align}
\begin{split}\label{geodesic1}
\left[ \frac{1}{2}\partial_s \bar h_{m n} - \partial_m \bar h_{sn} -  \frac{ \left(\bar h_{m n} F_{sr} - \tau_s \partial_r \bar h_{mn}\right) \dot x^r}{2N} - \frac{\dot N \bar h_{mn} \tau_s}{N^2} \right]\dot x^m \dot x^n &\hphantom{=} \\ + \frac{\dot N \bar h_{s n} \dot x^n - \tau_s \bar h_{mn} \ddot x^m x^n}{N} &= \bar h_{sn} \ddot x^n  \, ,
\end{split}
\end{align}
where we have defined $N \equiv \tau_p \dot x^p$. Contracting \eqref{geodesic1} with $h^{s t}$ give us the geodesic equation

\begin{align}
\begin{split}\label{geodesic}
\ddot x^t + \Gamma^t_{mn} \dot x^m \dot x^n &= \frac{\dot N}{N} \dot x^t - \frac{\bar h_{mn} F_{sr}h^{st}}{2N} \dot x^m \dot x^n \dot x^r \, .
\end{split}
\end{align}

We want to construct a solution of \eqref{geodesic} such that $x^m(0)=X^m_0$ and $x^\mu(1)= X^m_0 + l_s \bar Y^m$ where we can identify the vector $\dot x^m(0)= l_s Y^m$. The following expansion on $l_s$  satisfying the previously mentioned conditions can be constructed

\begin{align}
\begin{split}\label{geoExpansion}
x^m = X^m_0 + \lambda l_s Y^m + \frac{\lambda^2}{2} l^2_s Y^m_2 + \mathcal{O}(l^3_s) \, ,
\end{split}
\end{align}
substituting \eqref{geoExpansion} in \eqref{geodesic} it follows that 

\begin{align}
\begin{split}\label{geoDisc}
\left( Y^t_2 + \Gamma^t_{m n} Y^m Y^n  \right) &= \frac{\tau_n Y^n_2 Y^t + \partial_m \tau_n Y^m Y^n Y^t - \frac{1}{2} \bar h_{mn} h^{ts} F_{sr}  Y^r Y^m Y^n }{ \tau_p Y^p} \, ,
\end{split}
\end{align}
where all the geometric background functions are evaluated at $X^m_0$. Equation \eqref{geoDisc} has a solution of the form 

\begin{align}
Y^t_2 = - \Gamma^t_{m n} Y^m Y^n -  G^t_{mn} Y^m Y^n \, ,
\end{align}
with $G^t_{mn}$ a tensor satisfying

\begin{align}\label{GEquation}
\tau_{(r} G^t_{mn)} &= \tau_s G^s_{(mn} \delta^t_{r)} - \frac{1}{2} \bar h_{(mn}  F_{r)s} h^{st} \, ,
\end{align}
For \eqref{GEquation} to have a solution it is necessary to impose $F_{rs} h^{rt} h^{sw}=0$, obtaining $G^t_{mn}= \frac{1}{2} \bar h_{mn} a_s h^{st}$, meaning that the quantum field $\bar Y^m$ can be written in terms of the covariant vector $Y^m$ as 

\begin{align}
\begin{split}\label{covariantExpansion}
\bar Y^m &= Y^m - \frac{l_s}{2} \left( \Gamma^m_{rs} + \frac{1}{2}\bar h_{r s} a_n h^{mn} \right) Y^r Y^s + \mathcal{O}\left( l^2_s \right) \, .
\end{split}
\end{align}

\section{Tree level contributions from the Dilaton}\label{diltrapp}
In this appendix we will compute the tree level contributions to the beta functions. To this end we will need the contribution to the energy-momentum tensor coming from \eqref{dilac} and then compute its (classical) trace. Notice that the energy-momentum tensor will receive a contribution from this term even when the worldsheet is flat. The result is given by 
\begin{equation} \label{diltrace}
\left(- \frac{2 \pi}{\alpha '}\right)\gamma^{\alpha\beta}T^{Dil}_{\alpha\beta}=- \Box_{\sigma}\phi=-\Box_{\sigma} X^m \partial_m\phi-\gamma^{\alpha\beta}\partial_\alpha X^m\partial_\beta X^n\partial_m\partial_n\phi,
\end{equation}
where $\Box_\sigma$ is the d'Alembertian on the worldsheet, $\Box_\sigma=\gamma^{\alpha\beta}\partial_\alpha\partial_\beta$. To rewrite this in a useful way we need the equations of motion for the classical fields. These are found by varying the Lagrangian \eqref{PolyakovB}:
\begin{align}
0=&-\gamma^{\alpha\beta}\left(\partial_p \bar h_{mn} -2 \partial_m \bar h_{n p}\right)\partial_\alpha X^m\partial_\beta X^n+2\bar h_{mp}\Box_{\sigma}X^m-\epsilon^{\alpha\beta}\left(\partial_p \bar B_{mn }-2\partial_m\bar B_{pn}\right)\partial_\alpha X^m\partial_\beta X^n\notag\\
& +2\Delta \lambda^\alpha\partial_\alpha X^m \partial_{[m} \tau_{p]}+\tau_p\partial_\alpha\Delta\lambda^\alpha -2\Sigma\lambda^\alpha\partial_\alpha X^m\partial_{[m}\aleph_{p]} - \aleph_p \partial_\alpha\Sigma\lambda^\alpha  \\
0=&\ e^\alpha_-\partial_\alpha X^m\tau_m+e^\alpha_-\left(\partial_\alpha\eta+\partial_\alpha X^m \aleph_m\right)\\
0=&\ e^\alpha_+\partial_\alpha X^m\tau_m-e^\alpha_+\left(\partial_\alpha\eta+\partial_\alpha X^m \aleph_m\right)\\
0=&\ \partial_\alpha\Sigma\lambda^\alpha \label{partsl},
\end{align}
where 
\begin{equation}
\Delta \lambda^\beta \equiv \lambda_- e^\beta_+ - \lambda_+ e^\beta_-,\qquad \qquad \qquad \Sigma \lambda^\beta \equiv \lambda_- e^\beta_+ + \lambda_+ e^\beta_- \, .
\end{equation}
We now multiply the  first equation by $\frac{1}{2}h^{pr}$, the second equation by $e^\beta_+\partial_\beta$ and the third one by $e^\beta_-\partial_\beta$ to find
\begin{align}
-\Box_\sigma X^r = &\  \left(\Gamma^r_{mn}+\hat \upsilon^r \partial_m\tau_n \right)\partial_\alpha X^m\partial_\beta X^n\gamma^{\alpha\beta} -\frac{1}{2}h^{rp}H_{pmn}\partial_\alpha X^m\partial_\beta X^n\epsilon^{\alpha\beta}\notag\\
&+\hat \upsilon^r \tau_m \Box_\sigma X^m+\ h^{rp}\Delta\lambda^\alpha\partial_\alpha X^m \partial_{[m} \tau_{p]}  - h^{rp}\Sigma\lambda^\alpha\partial_\alpha X^m \partial_{[m} \aleph_{p]} \label{boxdil1}\\
\left(\tau_m+\aleph_m\right)\Box X^m=&\ e^\alpha_+ e^\beta_- \left(\partial_m \tau_n+\partial_m\aleph_n\right) \partial_\alpha X^m\partial_\beta X^n - \Box_\sigma\eta \label{boxdil2}\\
\left(\tau_m-\aleph_m\right)\Box X^m=&\ e^\alpha_+ e^\beta_- \left(\partial_n \tau_m-\partial_n\aleph_m\right) \partial_\alpha X^m\partial_\beta X^n +\Box_\sigma\eta\label{boxdil3}
\end{align}
where we have also used \eqref{partsl} to simplify \eqref{boxdil1}. By adding and subtracting \eqref{boxdil2} and \eqref{boxdil3} we find 
\begin{align}
\tau_m\Box_\sigma X^m&=e^\alpha_+ e^\beta_- \left(\partial_{(m} \tau_{n)}+\partial_{[m}\aleph_{n]}\right) \partial_\alpha X^\mu\partial_\beta X^\nu=-\left(\gamma^{\alpha\beta}\partial_{m} \tau_{n}+\epsilon^{\alpha\beta}\partial_{m}\aleph_{n}\right) \partial_\alpha X^m\partial_\beta X^n \label{boxdil4}\\
\aleph_m\Box_\sigma X^m&=-\left(\epsilon^{\alpha\beta}\partial_{m} \tau_{n}-\gamma^{\alpha\beta}\partial_{m}\aleph_{n}\right) \partial_\alpha X^m\partial_\beta X^n -\Box_\sigma \eta 
\end{align}
Substituting \eqref{boxdil4} in \eqref{boxdil1} we finally have
\begin{align} \label{boxfin}
-\Box_\sigma X^r=&\left(\Gamma^r_{mn}\gamma^{\alpha\beta}-\hat\upsilon^r\partial_m\aleph_n\epsilon^{\alpha\beta}-\frac{1}{2}h^{rp}H_{pmn}\epsilon^{\alpha\beta}\right)\partial_\alpha X^m\partial_\beta X^n \notag\\
&+h^{rp}\Delta\lambda^\alpha\partial_\alpha X^m \partial_{[m} \tau_{p]}  -h^{rp}\Sigma\lambda^\alpha\partial_\alpha X^m \partial_{[m} \aleph_{p]}.
\end{align}
Now that we have an expression for $\Box_\sigma X^r$ in terms of $\partial_\alpha X^r$ we can rewrite \eqref{diltrace} as
\begin{align}\label{dilatonContribution}
\gamma^{\alpha\beta}T^{Dil}_{\alpha\beta}= & -\frac{l_s^2}{2\pi} \left[- \gamma^{\alpha\beta} D_mD_n\phi - \epsilon^{\alpha\beta}\left(\hat\upsilon^r D_r\phi \, \partial_m\aleph_n+\frac{1}{2}h^{rp}D_r\phi H_{pmn}\right) \right]\partial_\alpha X^m\partial_\beta X^n\notag\\
&- \frac{l_s^2}{2\pi} h^{rp}D_r\phi\, \partial_{[m} \tau_{p]}\, \Delta\lambda^\alpha\partial_\alpha X^m  + \frac{l_s^2}{2\pi} h^{rp}D_r\phi \, \partial_{[m} \aleph_{p]}\, \Sigma\lambda^\alpha\partial_\alpha X^m \\ \notag =& - \frac{l_s^2}{4 \pi} \left[ \beta^\phi_{rs} \partial_\alpha X^r_0 \partial_\beta X^s_0 \gamma^{\alpha \beta} + \bar \beta^\phi_{rs }\partial_\alpha X^r_0 \partial_\beta X^s_0 \epsilon^{\alpha \beta} + \beta^\phi_m \Delta \lambda^\alpha \partial_\alpha X^m + \bar \beta^\phi_m \Sigma \lambda^\alpha \partial_\alpha X^m   \right],
\end{align}
from which one can easily read the dilaton contributions to the beta functions \eqref{beta1}-\eqref{beta6} :

\begin{align}
	\begin{split}\label{dilatonContributionsBeta}
		\beta^\phi_{mn} &=-2 \mathring{D}_{(m} \mathring{D}_{n)} \phi - 2G^t_{mn} \mathring{D}_t \phi\\
		\bar \beta^\phi_{mn} &= -\mathfrak{h}_{mn} \hat \upsilon^r \mathring{D}_r \phi - h^{rp}\mathring{D}_r \phi H_{pmn} \\  
		\beta^\phi_m &= h^{rp} \mathring{D}_r \phi F_{mp} \\ 
		\bar \beta^\phi_m &=- h^{rp} \mathring{D}_r \phi \mathfrak{h}_{mp}
	\end{split}
\end{align} 

For completeness we mention that the time projection of \eqref{boxdil1} is given by
\begin{align}
\partial_\alpha \Delta\lambda^\alpha&=\left[\hat\upsilon^p\left(D_p\bar h_{mn}-2D_m\bar h_{np}\right)\gamma^{\alpha \beta}+\left(\hat\upsilon^p H_{pmn}+2\Phi \mathfrak{h}_{mn}\right)\gamma^{\alpha \beta}\right]\partial_\alpha X_0^m\partial_\beta X_0^n\nonumber\\
&\hphantom{=}+a_m\Delta\lambda^\alpha\partial_\alpha X_0^m-\mathfrak{e}_m\Sigma\lambda^\alpha\partial_\alpha X_0^m.
\end{align}

\section{Covariant expansion of one loop effective action  }\label{appendixRecombine}

We will make use of \eqref{quantumFields}, \eqref{expansionX0} and \eqref{expansions} to write down the covariant expansion of the  couplings appearing on the Polyakov action \eqref{Polyakov}. Starting with the $\lambda-\eta$ coupling

\begin{align}\label{pullbackEta}
	\int \frac{ d^2 \sigma}{l_s^2} e \left[ \left(\lambda_+ e^\beta_- + \lambda_- e^\beta_+ \right) \partial_\beta \eta  \right] &= \int d^2 \sigma e \left[ \Sigma \bar \Lambda^\beta  \mathring{\nabla}_\beta \bar H\right]+ \mathcal{O}\left( l_s \right) \, ,
\end{align}
where we have defined $ \Sigma \bar \Lambda^\beta\equiv \left(\bar \Lambda_+ e^\beta_- + \bar \Lambda_- e^\beta_+ \right)$. We can then look at the $\bar h_{\alpha \beta}$ coupling 

\begin{align}
\begin{split}\label{pullbackH}
 \int \frac{ d^2 \sigma}{l_s^2}e \gamma^{\alpha \beta} \bar h_{\alpha \beta}(X) &= \int d^2 \sigma e  \left[  \bar h_{mn} \mathring{\nabla}_\alpha Y^m \mathring{\nabla}^\alpha Y^n \right] \\ &\hphantom{=}   +\int d^2 \sigma e \left[ 2 \left( \mathring{D}_s \bar h_{mn}\right) \mathring{\nabla}_\alpha Y^m Y^s \partial^\alpha X^n_0 \right] \\ 
&\hphantom{=} +   \int d^2 \sigma e \left[  \left( \frac{1}{2} \mathring{D}_r \mathring{D}_s \bar h_{mn} +  \mathring{R}^t_{\hphantom{t}rsm} \bar h_{nt} \right) Y^r Y^s \partial_\alpha X^m_0 \partial_\beta X^n_0 \right] \\ &\hphantom{=} + \mathcal{O}\left( l_s \right) \, ,
\end{split}
\end{align}
where we can recall that $H=d \bar B$, $F=d\tau$, $\mathfrak{h}=d\aleph$, $\Delta \lambda^\beta \equiv \lambda^0_- e^\beta_+ - \lambda^0_+ e^\beta_- $, and  $\Sigma \lambda^\beta \equiv \lambda^0_- e^\beta_+ + \lambda^0_+ e^\beta_-$. Moving into the vector couplings we have

\begin{align}
\begin{split}\label{pullbackTau}
 \int \frac{ d^2 \sigma}{l_s^2} e \left[\lambda_{\pm}e^\alpha_{\mp} \tau_\alpha(X)  \right] &=  \int d^2 \sigma e \left[ \bar \Lambda_{\pm} e^\alpha_{\mp}  \left(  \mathring{\nabla}_\alpha \left(\tau_m Y^m\right) - F_{mr} Y^r \partial_\alpha X^m_0  \right) \right]  \\ &\hphantom{=} +\int d^2 \sigma e \left[ \frac{1}{2} \left( F_{mn}  \right)  \lambda^0_\pm e^\alpha_{\mp} Y^m \mathring{\nabla}_\alpha Y^n \right] \\ &\hphantom{=} + \int d^2 \sigma e \left[\frac{1}{2} \left( \mathring{D}_r F_{sm}  \right)\vphantom{\frac{1}{2}} Y^r Y^s \lambda_{\pm}^0 e^\alpha_{\mp} \partial_\alpha X^m_0  \right] \\ &\hphantom{=} + \mathcal{O}\left( l_s \right) \, ,
\end{split}
\end{align}


\begin{align}
\begin{split}\label{pullbackAleph}
 \int\frac{ d^2 \sigma}{l_s^2} e \left[ \Sigma \lambda^\alpha \aleph_\alpha(X)  \right] &=  \int d^2 \sigma e \left[\Sigma \bar \Lambda^\alpha \left(  \aleph_m \mathring{\nabla}_\alpha Y^m +\mathring{D}_r \aleph_m Y^r \partial_\alpha X^m_0  \right) \right] \\ &\hphantom{=} + \int d^2 \sigma e \left[ \frac{1}{2} \left( \mathfrak{h}_{mn} \right) \Sigma \lambda^\alpha Y^m \mathring{\nabla}_\alpha Y^n \right]\\&\hphantom{=} + \int d^2 \sigma \left[ \frac{1}{2} \left(\mathring{D}_r \mathfrak{h}_{sm} \right) Y^r Y^s \Sigma \lambda^\alpha \partial_\alpha X^m_0 \right] \\ &\hphantom{=}  + \mathcal{O}\left( l_s \right) \, ,
\end{split}
\end{align}
where we have used \eqref{Dtauh} as well as the identity 

\begin{align}
\mathring{R}^t_{srm} \aleph_t = -\mathring{D}_r \mathring{D}_m \aleph_s +	 \mathring{D}_m \mathring{D}_r \aleph_s\, .
\end{align}

We can finally move to the last coupling 

\begin{align}
\begin{split}\label{pullbackB}
\int \frac{d^2\sigma\, e}{l_s^2}  \epsilon^{\alpha \beta} \bar B_{\alpha \beta}(X) &= \int d^2 \sigma e \epsilon^{\alpha \beta}\left[\left( H_{smn} \right) \mathring{\nabla}_\alpha Y^m Y^s \partial_\beta X^n_0   \right] \\ &\hphantom{=} + \int d^2 \sigma e \epsilon^{\alpha \beta} \left[ \frac{1}{2} \left(\mathring{D}_r H_{smn}\right) Y^r Y^s \partial_\alpha X^m_0 \partial_\beta X^n_0 \right] \\ &\hphantom{=}   + \mathcal{O}\left( l_s  \right) \, ,
\end{split}
\end{align}
where we have used the identity

\begin{align}
	\begin{split}
		\int d^2 \sigma e \epsilon^{\alpha \beta} \left[ \bar B_{mn} \mathring{\nabla}_\alpha Y^m \mathring{\nabla}_\beta Y^n\right] &= \int d^2 \sigma e \epsilon^{\alpha \beta} \left[\left( \mathring{D}_n \bar B_{sm}  \right) Y^s \mathring{\nabla}_\alpha Y^m \partial_\beta X^n_0 \right. \\ &\hphantom{=} \hphantom{\int d^2\sigma e \epsilon^{\alpha \beta}[[[[[} \left.+ \frac{1}{2} \left(\mathring{R}^t_{rmn} \bar B_{ts} \right) Y^r Y^s \partial_\alpha X^m \partial_\beta X^n \right]\,. 
	\end{split}
\end{align}

Before combining \eqref{pullbackEta}, \eqref{pullbackH}, \eqref{pullbackTau}, \eqref{pullbackAleph} and \eqref{pullbackB} to write down the action $\bar S_0$ we will take a look at the transformation properties of $\bar H$ inherited from the Kalb-Ramond U(1) transformation, namely if the transformation of the original fields are

\begin{align}
	\begin{split}
		\delta_\Lambda \aleph_m &= \partial_m \Lambda_u(X) \, , \\
		\delta_\Lambda \eta &= - \Lambda_u(X) \, ,
	\end{split}
\end{align} 
the quantum field $\bar H$ will transform as

\begin{align}
	\delta \bar H &= -D_r \Lambda_u Y^r + \mathcal{O}\left( l_s \right)\,.
\end{align}
It is then convenient to define a new field $\hat H$ as

\begin{align}
	\hat H = \bar H + \aleph_m Y^m
\end{align}
such that $\delta \hat H = \mathcal{O}\left(l_s\right)$, making it invariant under the Kalb-Ramond U(1) transformation at the level of the action $\bar S_0$. By making use of this field redefinition the action $\bar S_0$ is written as 

\begin{align}
\begin{split}
	\bar S_0 &= -  \int \frac{d^2\sigma e}{4\pi}\left[ \bar h_{mn} \mathring{\nabla}_\alpha Y^m \nabla^\alpha Y^n - \bar \Lambda_+ e^\beta_- \left(\mathring{\nabla}_\beta \hat H +  \mathring{\nabla}_\beta \left(\tau_m Y^m\right) \right)- \bar \Lambda_-e^\beta_+ \left(\mathring{\nabla}_\beta \hat H -  \mathring{\nabla}_\beta \left(\tau_m Y^m\right)\right) \right] \\
	&\hphantom{=} - \int \frac{d^2 \sigma e}{4 \pi} \left[ \bar \Lambda_+ Y^r \left(F_{mr} + \mathfrak{h}_{mr} \right) e^\beta_- \partial_\beta X^m_0 - \bar \Lambda_- Y^r \left( F_{mr} - \mathfrak{h}_{mr}  \right) e^\beta_+ \partial_\beta X^m_0 \right] \\
	&\hphantom{=} - \int \frac{d^2 \sigma e}{4 \pi}\left[ \left(\gamma^{\alpha \beta} A_{smn} + \epsilon^{\alpha \beta} \bar A_{smn} \right)  Y^s \mathring{\nabla}_\alpha Y^m \partial_\beta X^n_0 + \frac{1}{2} \left( \Delta \lambda^\beta F_{mn} - \Sigma \lambda^\beta \mathfrak{h}_{mn} \right) Y^m \mathring{\nabla}_\alpha Y^n  \right] \\
	&\hphantom{=} -\int \frac{d^2 \sigma e}{4 \pi} \left[ \left( \gamma^{\alpha \beta} C_{r s mn} + \epsilon^{\alpha \beta} \bar C_{rsmn} \right) Y^r Y^s \partial_\alpha X^m_0 \partial_\beta X^n_0 + \left( \Delta\lambda^\alpha B_{rsm} + \Sigma \lambda^\alpha \bar B_{rsm} \right) Y^r Y^s \partial_\alpha X^m_0 \right]\, ,\end{split}
\end{align}
where the coefficients $\{A,\bar A,C,\bar C,B,\Bar B \}$ are given by
\begin{align}
	\begin{split}\label{coefficientsAppendix}
		A_{smn}&=2\mathring{D}_s \bar h_{mn} \\
		\bar A_{smn} &=H_{smn} \\
		C_{rsmn} &=   \frac{1}{2} \mathring{D}_r \mathring{D}_s \bar h_{mn}  + \mathring{R}^t_{\hphantom{t}(rs)(m} \bar h_{n)t}  \\
		\bar C_{rsmn} &= \frac{1}{2} \mathring{D}_r H_{smn}  \\
		B_{rsm} &=\frac{1}{2}\mathring{D}_r F_{sm}  \\ 
		\bar B_{rsm} &=  - \frac{1}{2} \mathring{D}_r {\mathfrak{h}}_{sm}\, .
	\end{split}
\end{align}

\section{Beta functions derivation}\label{betaF}

Making use of decomposition \eqref{TildeDecomp} and assuming we are working on the critical spacetime dimension the Weyl variation of the effective action can be written as 

\begin{align}
	\begin{split}
		\delta_\psi \bar \Gamma[\Psi_0](0) &= \delta_\psi \langle  \mathcal{S}_1 + \mathcal{\tilde S}_1 + \mathcal{S}_2 + \mathcal{\tilde S}_2 \rangle_0 + \frac{i}{2} \delta_\psi \langle \left(\mathcal{\tilde S}_1 \mathcal{\tilde S}_1  + 2 \mathcal{\tilde S}_1 \mathcal{S}_1 \right) + \mathcal{S}_1 \mathcal{S}_1 \rangle_0 + \mathcal{O}\left( D^3 \right)
\end{split}
\end{align}
where we have also made use of the Ward identity  
\begin{align}\label{Ward1}
	\int d^2 \sigma J^\alpha_{IJ} \langle Y^I(\sigma) \partial_\alpha Y^J(\sigma) \rangle &= - \frac{1}{2} \int d^2 \sigma  \langle Y^I(\sigma) Y^J(\sigma) \rangle \partial_\alpha J^\alpha_{IJ}
\end{align}
with $J^\alpha_{IJ}$ an arbitrary spacetime tensor to move the disconnected part of the variation to one order higher in derivatives. We will start by computing the $\mathcal{S}$ two point correlations

\begin{align}
\begin{split}\label{2pCont}
		\delta_\psi \langle \mathcal{S}_1 + \mathcal{S}_2 \rangle &=  - \int \frac{d^2 \sigma \psi}{4 \pi} \left[ - \mathring{R}_{mn} + \ \left( \frac{1}{2} \mathring{D}\cdot a + a^2 \left(\frac{d_c}{4} + \frac{3}{4} \right)  \right) \bar h_{mn}  \right]   \gamma^{\alpha \beta}\partial_\alpha X^m_0 \partial_\beta X^n_0 \\
		&\hphantom{=} - \int \frac{d^2 \sigma \psi}{4 \pi} \left[ \frac{1}{2} h^{rs} \mathring{D}_r H_{smn} - \mathring{D}_m \mathfrak{e}_n + \left(\frac{d_c+2}{4}\right) a_r h^{rs} H_{smn} + \frac{\mathring{D}_t \hat \upsilon^t }{2} \mathfrak{h}_{mn} \right] \epsilon^{\alpha \beta}\partial_\alpha X^m_0 \partial_\beta X^n_0 \\
		&\hphantom{=} - \int \frac{d^2 \sigma \psi}{4 \pi} \left[\frac{1}{2} \mathring{D}\cdot a + \left(\frac{d_c}{4} + \frac{3}{4} \right)a^2  \right]  \tau_n \Delta \lambda^\alpha \partial_\alpha X^n_0\\
		&\hphantom{=} - \int \frac{d^2 \sigma \psi}{4 \pi} \left[ - \frac{1}{2} \mathring{D} \cdot \mathfrak{e} - \left( \frac{d_c}{4} + \frac{3}{4} \right) a \cdot \mathfrak{e} \right] \tau_n \Sigma \lambda^\alpha \partial_\alpha X^n_0 + ...
\end{split}
\end{align}
 where we have neglected terms that will not contribute to the final result, $\cdot$ denotes an inner product with respect to $h^{rs}$ and where we have used that 
\begin{align}
 \delta_\psi \int d^2 \sigma J_{IJ}	\langle Y^I(\sigma) Y^J(\sigma) \rangle &= \int d^2 \sigma J_{ij} \delta^{ij} \psi \, .  
\end{align}
with $J_{IJ}$ an arbitrary tensor, this last identity follows from the renormalization of the propagators \eqref{propagators}. In deriving \eqref{2pCont} we have introduced a total derivative \footnote{If this total derivative is not included then the U(1) mass variation of the antisymmetric beta function will not be zero but rather a total derivative, leaving the effective action invariant but not the beta function itself.}  $\int d^2 \sigma \partial_\alpha \mathfrak{e}^\alpha$, made use of the Ward identity \eqref{Ward1}, the background equation  \eqref{boxfin}, the Bianchi identity 

\begin{align}
	\begin{split}\label{Bianchi}
		\left( \mathring{D}_m \bar h_{sn} + \mathring{D}_n \bar h_{sm} - \mathring{D}_s \bar h_{mn} + \bar h_{mn} a_s \right) h^{ts} &= 0\,  ,
	\end{split}
\end{align}
and the TNC identities 

\begin{align}
	h^{rs} \left( \frac{1}{2} \mathring{D}_r \mathring{D}_s \bar h_{mn} - \mathring{D}_r \mathring{D}_{(m} \bar h_{n)s} \right) &= \frac{1}{2} \left(\mathring{D} \cdot a + a^2 \right) \bar h_{mn} + \mathring{D}_{(m} \bar h_{n)s} a_r h^{rs} \, , \\
	\mathring{D}_m \bar h_{ns} h^{rs} a_s &= \left( \Delta_T \right)^r_n \mathring{D}_m a_r - \frac{1}{2} a_m a_n- \frac{1}{2} a^2 \bar h_{mn} \, , \\
	h^{rs} \mathring{R}^t_{(mn)r} \bar h_{st} &= - \mathring{R}_{mn} - \frac{1}{4} a_m a_n - \frac{1}{2} \left( \Delta_S \right)^t_{(m} \mathring{D}_{n)}a_t \, , \\
	h^{rs} \hat \upsilon^t \mathring{D}_r \bar h_{st} &= - \mathring{D}_t \hat \upsilon^t \, ,
\end{align}
where $\left(\Delta_T\right)^r_m$ and $\left(\Delta_S\right)^r_m$ are the usual TNC temporal and spatial projectors 

\begin{align}
	\begin{split}
		\left(\Delta_T\right)^r_m &\equiv - \hat \upsilon^r \tau_m \, , \\ 
		\left(\Delta_S\right)^r_m &\equiv h^{rt} \bar h_{tm} \, .
	\end{split}
\end{align}

To compute the four point functions arising from $\langle \mathcal{S}_1\mathcal{S}_1 \rangle_0$ we will need the non-vanishing four point function variations identities 

\begin{align}
	\begin{split}\label{4pC}
		\delta_\psi \int d^2 \sigma  d^2 \sigma' J_{IJKL} \langle Y^I(\sigma) \partial_\alpha Y^J(\sigma) Y^K(\sigma') \partial_\beta Y^L(\sigma') &=  (-2 \pi i ) \int d^2 \sigma \psi J_{ijkl} \delta^{ik} \delta^{jl} \gamma_{\alpha \beta} \\
		\delta_\psi \int d^2 \sigma d^2 \sigma' J_{IJ} \langle \Lambda_+(\sigma) Y^I(\sigma) \Lambda_-(\sigma') Y^J(\sigma') \rangle &=\left( -4 \pi i \right) \int d^2 \sigma \psi J_{ij} \delta^{ij} \\
		\delta_\psi \int d^2 \sigma d^2 \sigma' J_{IJK} \langle Y^I(\sigma) e^\beta_{\mp} \partial_\beta Y^J(\sigma) \Lambda_\pm(\sigma') Y^K(\sigma') \rangle &= c_0\int d^2 \sigma \psi  \frac{J_{itj}\hat \upsilon^t \delta^{ij}}{\sqrt{2 \Phi}} + \mathcal{O}\left(D^3 \right) 
	\end{split}
\end{align} 

with $\{J_{IJKL},J_{IJ},J_{IJK}\}$ arbitrary $\mathcal{O}\left(D^2 \right)$ tensors, and $c_0$ an arbitrary constant \footnote{This identity can be derived through integration by parts and making use of propagators \eqref{propagators}.}. The presence of $c_0$ might seem like a problem to the uniqueness of the resulting beta functions, however by noting that the $U(1)$ mass symmetry is non-compatible with the derivative expansion\footnote{A $U(1)$ mass transformation changes the $\mathcal{O}\left(D\right)$ of the actions $S^{[a]}_0$.} we find that constants of the $c_0$ type will be completely fixed by asking for $U(1)$ mass invariance at second order in covariant derivatives. The ambiguity in defining the $\mathcal{O}\left(D\right)$ can also be seen from the two point function Ward identity \eqref{Ward1} as well as from the four point function identity

\begin{align}
	\begin{split}
		\int d^2 \sigma d^2 \sigma' \left(V^\alpha_{(IJ)} V^\beta_{(KL)} \right) \langle Y^I (\sigma) \partial_\alpha Y^J (\sigma) Y^K(\sigma') \partial_\beta Y^L(\sigma') \rangle &= \mathcal{O}\left(D^3 \right) \, ,
	\end{split}
\end{align}
with $V^\alpha_{IJ}$ an arbitrary tensor. Making use of \eqref{4pC} it is found that  

\begin{align}
	\begin{split}\label{S12}
		\frac{i}{2}\delta_\psi \langle \mathcal{S}_1 \mathcal{S}_1 \rangle &= - \int \frac{d^2 \sigma \psi }{4 \pi} \left[ \frac{1}{4} H_{rsm} H_{twn} h^{rt} h^{sw} + c_1 \left( \Delta_T\right)^t_m \mathring{D}_n a_r + c_4 \mathfrak{e}_r h^{rs} \left(\Delta_T \right)^t_m H_{nts} \right. \\ &\hphantom{= - \int \frac{d62 \sigma \psi}{4 \pi} } \left.\vphantom{\frac{1}{4}} + \frac{e^2 \left(2 \Phi \tau_m \tau_n + \bar h_{mn} \right) - e_m e_n}{2} - \bar h_{mn} \left( \mathfrak{e}^2 + \frac{a^2}{4} \right) \right. \\ &\hphantom{= - \int \frac{d62 \sigma \psi}{4 \pi} } \left.\vphantom{\frac{1}{4}} - a^2 \Phi \tau_m \tau_n\left( c_1 + \frac{5}{2} \right) \right]\gamma^{\alpha \beta} \partial_\alpha X^m_0 \partial_\beta X^n_0 \\
		&\hphantom{=} -  \int \frac{d^2 \sigma \psi }{4 \pi} \left[ - \frac{a_r h^{rs} H_{smn}}{2} + \left(c_3 - \frac{1}{2} \right) a_r h^{rs} \left(\Delta_T \right)^t_m H_{snt} \right. \\&\hphantom{= - \int \frac{d62 \sigma \psi}{4 \pi} } \left.\vphantom{\frac{1}{4}} + c_2 \left(\Delta_T \right)^r_m \mathring{D}_n \mathfrak{e}_r + a_m \mathfrak{e}_n \right] \epsilon^{\alpha \beta}\partial_\alpha X^m_0 \partial_\beta X^n_0 \\
		&\hphantom{=} - \int \frac{d^2 \sigma \psi}{4\pi} \left[ \left( c_5 a^2 + c_6 \mathfrak{e}^2\right)\Delta \lambda^\alpha + \left( c_7 a \cdot \mathfrak{e} \right)\Sigma \lambda^\alpha \right] \tau_n  \partial_\alpha X^n_0 
	\end{split}
\end{align}

To derive \eqref{S12} we made use of the projected Bianchi identity 

\begin{align}
	\begin{split}
		\left(\mathring{D}_r \bar h_{sn} - \mathring{D}_s \bar h_{rn} - \frac{1}{2} a_r \bar h_{sn} + \frac{1}{2} a_s \bar h_{rn} \right) &=0 \, ,
	\end{split}
\end{align}
the Bianchi identity \eqref{Bianchi}, the TNC identities 

\begin{align}
	\begin{split}
		\mathring{D}_m \bar h_{ns} h^{rs} a_s &= \left( \Delta_T \right)^r_n \mathring{D}_m a_r - \frac{1}{2}a_m a_n - \frac{1}{2} a^2 \bar h_{mn} \, ,\\
		\mathring{D}_m \bar h_{ns} h^{rs} \mathfrak{e}_s &= \left( \Delta_T \right)^r_n \mathring{D}_m \mathfrak{e}_r - \frac{1}{2}\mathfrak{e}_m a_n - \frac{1}{2} \left(\mathfrak{e} \cdot a \right)\bar h_{mn} \, ,
	\end{split}
\end{align}
and where we have used the Ward identity \eqref{4pC} to introduce the  $\mathcal{O}\left(D^2\right)$ zeros
\begin{align}
		\int \frac{d^2 \sigma}{4 \pi} \left[ \left(\frac{a_m a_n -\mathfrak{e}_m \mathfrak{e}_n + \left(a^2 - \mathfrak{e}^2 \right) \left(\bar h_{mn} + 2 \Phi \tau_m \tau_n \right)}{2}  \right) \gamma^{\alpha \beta} \right]\partial_\alpha X^m_0 \partial_\beta X^n_0 &\hphantom{=} \\ \nonumber + \int \frac{d^2 \sigma}{4\pi}\left[ \left( a_m \mathfrak{e}_n \right) \epsilon^{\alpha \beta} \right]\partial_\alpha X^m_0 \partial_\beta X^n_0&= \mathcal{O}\left( D^3 \right) \, ,\\
		\int \frac{d^2 \sigma}{4\pi} \left[ a_m a_n + a^2 \left(\bar h_{mn} + 2 \Phi \tau_m \tau_n \right) \right] \partial_\alpha X^m_0 \partial^\alpha X^n_0 &= \mathcal{O}\left( D^3 \right) \, .
\end{align}

Following an analogous procedure we can compute the contributions from $\mathcal{\tilde S}$, in particular we find that 

\begin{align}
	\begin{split}\label{STilde}
		\delta_\psi \left\langle \mathcal{\tilde S}_1  + \mathcal{\tilde S}_2 + \frac{i}{2} \mathcal{\tilde S}_1 \left(\mathcal{\tilde S}_1 + \mathcal{S}_1 \right) \right\rangle &=  - \int \frac{d^2 \sigma \psi}{4 \pi} \left[ \frac{\left(\Delta_T \right)^t_m \mathring{D}_n a_t}{2}   + \tilde c_0 \Phi \left(a \cdot \mathfrak{e} \right) \tau_m \tau_n\right. \\ &\hphantom{= - \int \frac{d^2 \sigma \psi}{4 \pi}} \left.\vphantom{\frac{1}{2}} + \left(c_0 -1 \right)a^2 \Phi \tau_m \tau_n \right] \partial_\alpha X^m_0 \partial^\alpha X^n_0 \\
		&\hphantom{=} - \int \frac{d^2 \sigma \psi}{4\pi} \left[ -\frac{a_r h^{rs} \left(\Delta_T\right)^t_m H_{snt}}{2} \right] \epsilon^{\alpha \beta}\partial_\alpha X^m_0 \partial_\beta X^n_0 \\
		&\hphantom{=} - \int \frac{d^2 \sigma \psi}{4\pi}\left[ \frac{a \cdot e }{4} \Sigma \lambda^\alpha - \frac{a^2}{4}\Delta \lambda^\alpha \right] \tau_n \partial_\alpha X^n_0 \, ,
	\end{split}
\end{align}
where we have used the identities
 
\begin{align}
	\begin{split}
		\left(\bar h_{rs} - \bar h_{rp} \bar h_{sq} h^{pq} \right) \mathring{D}_m e^r_i \mathring{D}_n e^s_j \delta^{ij} &= - \frac{1}{2}a^2 \tau_m \tau_n \Phi \, , \\
		\left(\delta^t_s - h^{tq}\bar h_{qs} \right) e^r_i \mathring{D}_m e^s_j \delta^{ij} &=  \frac{1}{2} a_s h^{rs}\left( \Delta_T \right)_m^t \, ,
	\end{split}
\end{align}
We can note that the analogous $\mathcal{\tilde S}$ computation in the standard bosonic string will result in a vanishing result, however in our case this is no longer true as $h^{rs} \bar h_{st} \neq \delta^r_t$  as well as due to the presence of a non-trivial coupling with the Lagrange multipliers. Combining \eqref{2pCont}, \eqref{S12}, \eqref{STilde}, and the classical dilaton contribution \eqref{dilatonContributionsBeta} results in the beta functions 

\begin{align}\label{beta1A}
 \beta_m &= \left[\frac{1}{2} \mathring{D}\cdot a + \left( \frac{d_c}{4} + \frac{1}{2} + c_5 \right) a^2 + c_6 \mathfrak{e}^2 - a \cdot \mathring{D}\phi \right] \tau_m \, ,\\
 \label{beta2A}\bar \beta_m &= - \left[ \frac{1}{2} \mathring{D}\cdot \mathfrak{e} + \left(\frac{d_c}{4} + \frac{1}{2} - c_7 \right) a \cdot \mathfrak{e} - \mathfrak{e} \cdot \mathring{D}\phi  \right] \tau_m \, , \\
 \label{beta3A}\beta_{mn} &= - \mathring{R}_{mn} + \frac{1}{4}H_{rsm} H_{twn} h^{rt} h^{sw} + \left[ \frac{1}{2} \mathring{D} \cdot a + \left( \frac{d_c}{4} + \frac{1}{2} \right) a^2 - \mathfrak{e}^2 - a \cdot \mathring{D}\phi   \right] \bar h_{mn} \\ \nonumber
 &\hphantom{=} + \left[c_1 + \frac{1}{2}\right] \left(\Delta_T\right)^t_m \mathring{D}_n a_t + c_4 \mathfrak{e}_r h^{rs} \left( \Delta_T \right)^t_m H_{nts} + \left[ \tilde c_0 a \cdot \mathfrak{e} + \left(c_0 - c_1 - \frac{7}{2} \right) a^2 \right] \Phi \tau_m \tau_n \\ \nonumber
 &\hphantom{=} + \frac{\mathfrak{e}^2 \left(2 \Phi \tau_m \tau_n + \bar h_{mn} \right) - \mathfrak{e}_m \mathfrak{e}_n}{2} - 2 \mathring{D}_m \mathring{D}_n \phi \, , \\
 \label{beta4A}\bar \beta_{mn} &= \frac{1}{2} h^{rs} \mathring{D}_r H_{smn} + \frac{d_c}{4} a_r h^{rs}H_{smn}  - \mathring{D}_m \mathfrak{e}_n + c_2 \left( \Delta_T\right)^r_m \mathring{D}_n \mathfrak{e}_r  +a_m \mathfrak{e}_n + \frac{\mathring{D}_t \upsilon^t}{2} \mathfrak{h}_{mn}
 \\ \nonumber &\hphantom{=} + \left(c_3 -1 \right) a_r h^{rs} \left(\Delta_T \right)^t_m H_{snt} - \left(\hat \upsilon^r \mathfrak{h}_{mn} + h^{rp} H_{p m n} \right) \mathring{D}_r \phi\, .
 \end{align}
The free coefficients in \eqref{beta1A}-\eqref{beta4A} can be fixed by asking for $\{\beta,\bar \beta_m, \beta_{mn}, \bar \beta_{mn} \}$ to be gauge invariant, this condition fixes the coefficients to  $c_0=3,\tilde c_0=0,c_1 =-\frac{1}{2}, c_2 =2,c_3=1,$ $c_4=-1,c_5=0,c_6=-1,c_7=\frac{1}{2}$ resulting in the beta functions \eqref{beta1M}-\eqref{beta4M} presented in the main text.

\section{Fadeev-Popov gauge fixing}\label{ghosts}

The gauge symmetries of the theory are 

\begin{align}
\begin{split}
\delta e^\beta_{\pm} &= - \omega_{\pm} e^\beta_{\pm} + \xi^\mu \partial_\mu e^\beta_{\pm} - e^\mu_{\pm}\partial_\mu \xi^\beta \\
\delta \lambda_{\pm} &= - \omega_{\pm} \lambda_{\pm} + \xi^\mu \partial_\mu \lambda_{\pm}
\end{split}
\end{align}
with $\omega_{\pm}$ parametrizing local worldsheet Weyl/Lorentz transformations and $\xi^\mu$ parametrizing worldsheet diffeomorphisms. Following the Faddeev-Popov procedure we can first compute the Faddeev-Popov determinant 

\begin{align}
\begin{split} \label{ghost0}
\Delta_{FP} &= \int Da D\bar aDb^* D \bar b^* Dc D d_+ D d_-  \left[ e^{ i S_{FP}} \right] \\
S_{FP} &=  \int \frac{d^2 \sigma e}{2 \pi l_s^2} \left[ b^*_\beta \left(c^\alpha \partial_\alpha e^\beta_+ - e^\alpha_+ \partial_\alpha c^\beta - d_+ e^\beta_+ \right) + \bar b^*_\beta \left(c^\alpha \partial_\alpha e^\beta_- - e^\alpha_- \partial_\alpha c^\beta - d_- e^\beta_- \right)  \right. \\
 &\hphantom{=} \hphantom{i \int d^2 \sigma aa}\left. + a_\alpha \left(\hat e^\alpha_+ - e^\alpha_+ \right) + \bar a_\alpha \left(\hat e^\alpha_- - e^\alpha_- \right) \vphantom{\bar b^*_\beta \left(c^\alpha \partial_\alpha e^\beta_- - e^\alpha_- \partial_\alpha c^\beta - c_+ e^\beta_- \right) }\right]
\end{split} 
\end{align}
where $\{c,d_{\pm},b^*,\bar b^* \}$ are Faddeev-Popov ghosts and anti-ghosts and $\{a_\mu,\bar a_\mu\}$ are bosonic Lagrange multipliers enforcing the gauge condition $\hat e$. The remaining BRST symmetry of the theory is given by 

\begin{align}
s X^m &= c^\alpha \partial_\alpha X^m, \quad \quad s \eta = c^\alpha \partial_\alpha \eta \\ 
s e^\beta_{\pm} &= c^\alpha \partial_\alpha e^\beta_{\pm} - e^\alpha_{\pm} \partial_\alpha c^\beta - d_{\pm} e^\beta_{\pm}\\
s \lambda_{\pm} &= - d_{\pm} \lambda_{\pm} + c^\alpha \partial_\alpha \lambda_{\pm} \\
s c^\beta &= c^\alpha \partial_\alpha c^\beta \\
s c^\beta_{\pm} &= c^\alpha \partial_\alpha c^\beta_{\pm} \\
s b^{*}_\beta &= a_\beta, \quad s a_\beta =0\\
s \bar b^*_\beta &= \bar a_\beta, \quad s \bar a_\beta = 0
\end{align}
Integrating over $\{a,\bar a, d_{\pm} \}$ imposes the constraints

\begin{align}
\begin{split}
e^\alpha_{\pm} &= \hat e^\alpha_\pm \\
b^*_\beta e^\beta_+ &= 0 \\
\bar b^*_\beta e^\beta_- &= 0
\end{split}
\end{align}
and the action simplifies into 

\begin{align}
S_{FP} =  \int\frac{d^2 \sigma e}{2 \pi l_s^2} \left[ c^\alpha e^\beta_+ \partial_\alpha b^*_\beta - b^*_\beta e^\alpha_+ \partial_\alpha c^\beta + c^\alpha e^\beta_- \partial_\alpha \bar b^*_\beta - \bar b^*_\beta e^\alpha_- \partial_\alpha c^\beta  \right] \equiv \int\frac{d^2 \sigma e}{2 \pi l_s^2}  \mathcal{L}_{FP}
\end{align}
where we have omitted the hat on the vielbeins for simplicity. By considering the action \eqref{ghost0} (after gauge-fixing), it is now possible to define the ghost energy momentum one form as in \eqref{momentumOne}

\begin{align}
\begin{split}\label{momentum2}
\tau^c_{\gamma \text{\tiny{FP}} } &= - \frac{2 \pi l_s^2}{e} \frac{\delta S_{FP}}{\delta e^\gamma_c} \\
&= \mathcal{L}_{FP} e^a_\gamma +\left[ \left(\delta^c_0 + \delta^c_1 \right) \left( b^*_\gamma \partial_- c^- + \partial_\alpha b^*_\gamma\, c^\alpha +b^*_-\partial_\gamma c^- \right) \right. \\ &\hphantom{=} \left.  +  \left(\delta^c_0 - \delta^c_1 \right) \left(\bar b^*_\gamma \partial_+ c^+ + \partial_\alpha \bar b^*_\gamma\, c^\alpha +\bar b^*_+\partial_\gamma c^+ \right)\right]
\end{split}
\end{align}
where we anticipated going to conformal gauge, i.e. the vielbeins are constant.

It is important to note that the anti-ghosts $\{b^*,\bar b^*\}$ will not be neutral under local Weyl transformations, meaning that we will need to supplement the theory with the transformation 

\begin{align}
\begin{split}
b^*_\beta \rightarrow f_{-} b^*_\beta, \quad \quad \bar b^*_\beta \rightarrow f_+ \bar b^*_\beta 
\end{split}
\end{align}
this implies the full condition for Weyl invariance is not \eqref{traceless} but rather

\begin{align}
\begin{split}
\left \langle  e^\gamma_c \bar \tau^c_\gamma +  C^+ \lambda_+ +  C^- \lambda_-  + b^*_\beta  B^\beta + \bar b^*_\beta  \bar B^\beta   \right \rangle &=0,
\end{split}
\end{align}
where $\bar \tau^c_\gamma = \tau^c_\gamma + \tau^c_{\gamma \text{\tiny{FP}} }  $ is the total energy momentum one form and $\{B^\beta, \bar B^\beta \}$ are the equations of motion for the anti-ghosts defined as 

\begin{align}\label{ghostConstraints}
B^\beta &=  - \frac{2 \pi \alpha}{e} \frac{\delta S_{FP}}{\delta b^*_\beta}, \quad \quad \bar B^\beta = - \frac{2 \pi \alpha}{e} \frac{\delta S_{FP}}{\delta \bar b^*_\beta}
\end{align}

In conformal gauge the ghost action takes the dimensionally extended form 

\begin{align}
\begin{split}
S_{FP}= -\int \frac{d^n \sigma}{2\pi}  \left[ \bar b \partial_+  \bar c +  \bar b \partial_-  c  \right] e^{(n-1)\rho}
\end{split}
\end{align}
where we have defined $\{ \bar b \equiv b^*_-, \bar c \equiv c^- , b \equiv \bar b^*_+ , c  \equiv c^+ \} $ and we have rescaled the ghosts such that the normalizaion of the action is $-1/2\pi$. The non vanishing real space propagators are given by 

\begin{align} \label{ghostsprop}
\begin{split}
\left \langle b(\sigma) c(\sigma') \right \rangle & = \frac{2 e^{-\rho}\Delta_2}{\left(\sigma - \sigma' \right)_+} \\
\left \langle \bar b(\sigma) \bar c(\sigma') \right \rangle &= \frac{2 e^{-\rho} \Delta_2}{\left(\sigma - \sigma' \right)_-}
\end{split}
\end{align}
where $\Delta_2$ is an overall factor that will not play any role in our results. To find these propagators we used the identity
\begin{equation}
\partial^2 \log\left(|\Delta \sigma|^2\right)=4\pi\,\delta(\Delta \sigma)\,.
\end{equation}

\section{Critical Dimension}\label{criticalDimension}
The condition for local Weyl invariance for the system   $Z_0 + Z_{FP}$ as described in \eqref{variationExplicit} is given by 

\begin{align}\label{WeylAppendix}
\langle \hat T \rangle \equiv \langle T^\gamma_{\gamma} + T^\gamma_{\gamma \text{\tiny{FP}}} + C^+ \Lambda_+ + C^- \Lambda_- + B^\beta b^*_\beta + \bar B^\beta \bar b^*_\beta \rangle = 0
\end{align}
where we have defined $T^\gamma_\delta \equiv e_c^\gamma \tau^c_\delta$ and $T^\gamma_{\delta \text{\tiny FP}} \equiv e^c_\gamma \tau^c_{\delta \text{\tiny{FP}}}$ and where we should do the substitution  $\{X,\lambda, \eta  \} \rightarrow \{Y,\Lambda, H \}$ in the energy momentum one forms given by \eqref{momentumOne} and \eqref{momentum2} as well as in the constraints \eqref{constraintsOne} and \eqref{ghostConstraints}. To analyze \eqref{WeylAppendix} we can use the following identity

\begin{align}
\int d^2 \sigma \delta_\psi \langle \hat T(\sigma) \rangle = \int d^2 \sigma \int d^2 \sigma' \psi(\sigma') \langle \hat T(\sigma) \hat T(\sigma')\rangle \, .
\end{align}

It is now our goal to compute the two point function of  traces $\hat T$. We can now note that in conformal gauge the  following holds 

\begin{align}
\begin{split}\label{traceSquare1}
\int d^2 \sigma d^2 \sigma' \langle \hat T\hat T' \rangle &= \int d^2 \sigma d^2 \sigma'  \left \langle \left( \vphantom{T^{'+}_+} T^+_+ + T^-_- + C^+ \Lambda_+  + C^- \Lambda_-   \right) \left( T^{'+}_+ + T^{'-}_- + C^{'+} \Lambda'_+  + C^{'-}\Lambda'_-  \right) \right \rangle  \\
&\hphantom{=} + \int d^2 \sigma d^2 \sigma' \left \langle       \left( \vphantom{T^{'+}_{+ \text{\tiny{FP}}}} T^+_{+ \text{\tiny{FP}}} + T^-_{- \text{\tiny {FP}}} + \bar b B^-  + b \bar B^+  \right) \left( T^{'+}_{+ \text{\tiny{FP}}} + T^{'-}_{- \text{\tiny {FP}}} + \bar b' B^{'-}  + b' \bar B^{'+}   \right) \right \rangle \\
& = \int d^2 \sigma d^2 \sigma'  \left[ \left \langle T^+_+ T^{'+}_{\hphantom{'}+} \right \rangle + 2 \left \langle T^+_+ T^{'-}_{\hphantom{'}-} \right \rangle + \left \langle T^-_- T^{'-}_{\hphantom{'}-} \right \rangle  +\left \langle T^+_{+\text{\tiny FP}} T^{'+}_{\hphantom{'}+ \text{\tiny FP}} \right \rangle  \right. \\ &\hphantom{=} \left. + 2 \left \langle T^+_{+\text{\tiny FP}} T^{'-}_{\hphantom{'}-\text{\tiny FP}} \right \rangle + \left \langle T^-_{-\text{\tiny FP}} T^{'-}_{\hphantom{'}- \text{\tiny FP}} \right \rangle       - \left \langle C^+ \Lambda_+ C^{'+} \Lambda'_+ \right \rangle - 2\left \langle C^+ \Lambda_+ C^{'-} \Lambda'_- \right \rangle   \right. \\ &\hphantom{=} \left. - \left \langle C^- \Lambda_- C^{'-} \Lambda'_- \right \rangle - \left \langle \bar b B^- \bar b' B^{'-}  \right \rangle - 2 \left \langle \bar b B^{-} b \bar B^{+} \right \rangle - \left \langle b \bar B^{+}  b' \bar B^{'+} \right \rangle   \right]
\end{split}
\end{align}
where we have denoted the dependence on $\sigma'$ by priming the variable itself and where 

\begin{align}
\begin{split}
T^{\pm}_{\pm} &= \frac{a e^\rho }{2} \lambda_{\pm} \left(\partial_\mp  \eta \pm  \partial_\mp  Y^0\right) \\
T^+_- &= - 2  \eta_{m n} \partial_- Y^m \partial_- Y^n  - \frac{a e^\rho}{2} \Lambda_- \left(\partial_- \eta -  \partial_- Y^0 \right)  \\
T^-_+ &=  - 2  \eta_{m n} \partial_+ Y^m \partial_+ Y^n  - \frac{a e^\rho}{2} \Lambda_+ \left(\partial_+ \eta +  \partial_+ Y^0 \right)  \\
T^{+}_{+ \text{\tiny{FP}}} &=- a e^\rho b \partial_{-} c \quad \quad T^{-}_{- \text{\tiny{FP}}} = - a e^\rho \bar b \partial_+ \bar c \\
T^{+}_{- \text{\tiny{FP}}} &= a e^\rho \left(\partial_+\bar b\, \bar c+2\bar b \partial_- \bar c + \partial_- \bar b\, \bar c \right)  \\
T^-_{+ \text{\tiny{FP}}} &= a e^\rho \left(\partial_- b\, c+2 b \partial_+ c + \partial_+  b\,  c \right) \\ 
C^{\pm} &= - \frac{a e^\rho }{2} \left( \partial_{\mp} \eta \pm \partial_{\mp} Y^0 \right) \\
B^-&= a e^\rho \partial_+ \bar c \\
\bar B^+ &= a e^\rho  \partial_- c\, .
\end{split}
\end{align}
To compute these correlators we will need the following real space propagators (that can be read from \eqref{propmat}):
\begin{align}
\begin{split}
\left \langle Y^i\left(\sigma\right) Y^j\left(\sigma'\right) \right \rangle &= \delta^{ij} \Delta_2 \ln \left( | \Delta \sigma |^2 \right) \\
\left \langle Y^0\left(\sigma \right) \Lambda_{\pm}\left(\sigma' \right) \right \rangle &= \frac{\mp 2 e^{-\rho}\Delta_2}{\left(\sigma - \sigma' \right)_\pm} \\
\left \langle H \left(\sigma \right) \Lambda_{\pm}\left(\sigma' \right) \right \rangle &=  \frac{- 2 e^{-\rho}\Delta_2}{\left(\sigma- \sigma' \right)_\pm}  \\
\
\left \langle \Lambda_{\pm}\left(\sigma \right) \Lambda_{\pm}\left(\sigma' \right) \right \rangle &= \frac{4 e^{-2 \rho} \Delta_2}{\left(\sigma - \sigma' \right)^2_\pm} \\
\left \langle \Lambda_{+}\left(\sigma\right) \Lambda_{-}\left(\sigma'\right) \right \rangle &=-4\pi\Delta_2 e^{-2 \rho} \delta(\sigma- \sigma' )
\end{split}
\end{align}
where $\Delta_2$ is an unimportant overall factor which was introduced after \eqref{ghostsprop}.   The contribution to the two point function from the constraints can then be computed to be

\begin{align}
\begin{split}\label{twoPointConstraints}
\left \langle C^+ \Lambda_+ C^{'+} \Lambda'_{+} \right \rangle &= - 16 \Delta^2_2 \partial_- \left(\frac{1}{\Delta \sigma_+} \right) \partial'_- \left( \frac{1}{\Delta \sigma_+}  \right) \\
\left \langle C^+ \Lambda_+ C^{'-} \Lambda'_{-} \right \rangle &= 0 \\
\left \langle C^- \Lambda_- C^{'-} \Lambda'_{-} \right \rangle &=- 16 \Delta^2_2 \partial_+ \left(\frac{1}{\Delta \sigma_-} \right) \partial'_+ \left(\frac{1}{\Delta \sigma_-}  \right) \\
\left \langle \bar b B^- \bar b' B^{'-} \right \rangle &= 16 \Delta^2_2  \partial_+ \left(\frac{1}{\Delta \sigma_-} \right) \partial'_+ \left(\frac{1}{\Delta \sigma_-}  \right) \\ 
\left \langle \bar b B^-  b' \bar B^{'+} \right \rangle &=0\\
\left \langle  b B^+  b' B^{'+} \right \rangle &= 16 \Delta^2_2 \partial_- \left(\frac{1}{\Delta \sigma_+} \right) \partial'_- \left( \frac{1}{\Delta \sigma_+} \,. \right)
\end{split}
\end{align}
Notice that the sum of these contributions is zero. From \eqref{twoPointConstraints} then we see that \eqref{traceSquare1} reduces to 

\begin{align}
\begin{split}
\int d^2 \sigma d^2 \sigma' \langle \hat T\hat T' \rangle &=  \int d^2 \sigma d^2 \sigma'  \left[ \left \langle T^+_+ T^{'+}_{\hphantom{'}+} \right \rangle + 2 \left \langle T^+_+ T^{'-}_{\hphantom{'}-} \right \rangle + \left \langle T^-_- T^{'-}_{\hphantom{'}-} \right \rangle   \right. \\ &\hphantom{=} \hphantom{ \int d^2 \sigma d^2 \sigma'  } \left.  +\left \langle T^+_{+\text{\tiny FP}} T^{'+}_{\hphantom{'}+ \text{\tiny FP}} \right \rangle  + 2 \left \langle T^+_{+\text{\tiny FP}} T^{'-}_{\hphantom{'}-\text{\tiny FP}} \right \rangle + \left \langle T^-_{-\text{\tiny FP}} T^{'-}_{\hphantom{'}- \text{\tiny FP}} \right \rangle   \right]
\end{split}
\end{align}
To further compute this in a consistent way we will need the conservation equation for $T^\delta_\gamma$ 

\begin{align}
\begin{split}
\partial_\delta T^\delta_\gamma &= 0  \\
\partial_- T^-_- &= -\partial_+ T^+ _ -\\
\partial_+ T^+_+ &= -\partial_- T^-_+
\end{split}
\end{align}
We can then note

\begin{align}
\begin{split}
\left \langle T^+_+ T^{'+}_+ \right \rangle &= \frac{\Delta_3 \Delta^2_2 \left(4d + 8 \right)}{3} \partial'_+ \partial'_- \delta(\sigma - \sigma') \\ 
\left \langle T^+_+ T^{'-}_- \right \rangle &= 0\\
\left \langle T^-_- T^{'-}_- \right \rangle &= \frac{\Delta_3 \Delta^2_2 \left(4 d + 8\right)}{3} \partial'_+ \partial'_- \delta(\sigma - \sigma') \\
\left \langle T^+_{+\text{\tiny FP}} T^{'+}_{+\text{\tiny FP}} \right \rangle &=  \frac{\Delta_3 \Delta^2_2 \left(- 104 \right)}{3} \partial'_+ \partial'_- \delta(\sigma - \sigma') \\
\left \langle T^+_{+\text{\tiny FP}} T^{'-}_{-\text{\tiny FP}} \right \rangle &= 0  \\
\left \langle T^-_{-\text{\tiny FP}} T^{'-}_{-\text{\tiny FP}} \right \rangle &=  \frac{\Delta_3 \Delta^2_2 \left(-104 \right)}{3} \partial'_+ \partial'_- \delta(\sigma - \sigma')
\end{split}
\end{align}
where $\Delta_3$ is another overall factor that does not change the final result. We can finally see that the central charge vanishes when $d=24$ and hence the critical dimension of TNC spacetime is $D=d+1=25$. 

\bibliographystyle{ieeetr} 
\bibliography{NCbib}

\begin{thebibliography}{10}

\bibitem{Cartan1923}
E.~Cartan, ``Sur les vari\'et\'es \`a connexion affine, et la th\'eorie de la
  relativit\'e g\'en\'eralis\'ee (premi\`ere partie),'' {\em Annales
  scientifiques de l'\'Ecole Normale Sup\'erieure}, vol.~40, pp.~325--412,
  1923.

\bibitem{Cartan2}
E.~Cartan, ``Sur les vari\'et\'es \`a connexion affine, et la th\'eorie de la
  relativit\'e g\'en\'eralis\'ee (premi\`ere partie) (suite),'' {\em Annales
  scientifiques de l'\'Ecole Normale Sup\'erieure}, vol.~3e s{\'e}rie, 41,
  pp.~1--25, 1924.

\bibitem{Friedrichs}
K.~Friedrichs, ``Eine invariante formulierung des newtonschen
  gravitationsgesetzes und des grenzberganges vom einsteinschen zum newtonschen
  gesetz,'' {\em Matematische Annalen}, vol.~98, pp.~566--575, 1928.

\bibitem{Inonu1952}
E.~Inönü and E.~P.~Wigner, ``Representations of the galilei group,'' {\em Il
  Nuovo Cimento Series 9}, vol.~9, pp.~705--718, 08 1952.

\bibitem{Bargmann:1954gh}
V.~Bargmann, ``{On Unitary ray representations of continuous groups},'' {\em
  Annals Math.}, vol.~59, pp.~1--46, 1954.

\bibitem{LevyLeblond:1967zz}
J.-M. Levy-Leblond, ``{Nonrelativistic particles and wave equations},'' {\em
  Commun. Math. Phys.}, vol.~6, pp.~286--311, 1967.

\bibitem{LeblondBook1}
J.-M. Levy-Leblond, {\em Galilei Group and Galilean Invariance}, pp.~221--299.
\newblock 12 1971.

\bibitem{Andringa:2010it}
R.~Andringa, E.~Bergshoeff, S.~Panda, and M.~de~Roo, ``{Newtonian Gravity and
  the Bargmann Algebra},'' {\em Class. Quant. Grav.}, vol.~28, p.~105011, 2011.

\bibitem{Niels2014}
M.~H. Christensen, J.~Hartong, N.~A. Obers, and B.~Rollier, ``Torsional
  newton-cartan geometry and lifshitz holography,'' {\em Phys. Rev. D},
  vol.~89, p.~061901, Mar 2014.

\bibitem{connection1}
K.~Jensen, ``{On the coupling of Galilean-invariant field theories to curved
  spacetime},'' {\em SciPost Phys.}, vol.~5, p.~11, 2018.

\bibitem{Son}
D.~Son, ``Newton-cartan geometry and the quantum hall effect,'' {\em
  arXiv:1306.0638 [cond-mat.mes-hall]}, 2013.

\bibitem{Hartong2014}
M.~H. Christensen, J.~Hartong, N.~A. Obers, and B.~Rollier, ``Torsional
  newton-cartan geometry and lifshitz holography,'' {\em Phys. Rev. D},
  vol.~89, p.~061901, Mar 2014.

\bibitem{Dieter}
D.~V. den Bleeken, ``Torsional newton{\textendash}cartan gravity from the large
  c expansion of general relativity,'' {\em Classical and Quantum Gravity},
  vol.~34, p.~185004, aug 2017.

\bibitem{Niels2018}
T.~Harmark, J.~Hartong, L.~Menculini, N.~A. Obers, and Z.~Yan, ``Strings with
  non-relativistic conformal symmetry and limits of the ads/cft
  correspondence,'' {\em Journal of High Energy Physics}, vol.~2018, p.~190,
  Nov 2018.

\bibitem{Callan}
C.~Callan, D.~Friedan, E.~Martinec, and M.~Perry, ``Strings in background
  fields,'' {\em Nuclear Physics B}, vol.~262, no.~4, pp.~593 -- 609, 1985.

\bibitem{Harmark2014}
T.~Harmark and M.~Orselli, ``Spin matrix theory: a quantum mechanical model of
  the ads/cft correspondence,'' {\em Journal of High Energy Physics},
  vol.~2014, p.~134, Nov 2014.

\bibitem{Niels2017}
T.~Harmark, J.~Hartong, and N.~A. Obers, ``Nonrelativistic strings and limits
  of the ads/cft correspondence,'' {\em Phys. Rev. D}, vol.~96, p.~086019, Oct
  2017.

\bibitem{Andringa:2012uz}
R.~Andringa, E.~Bergshoeff, J.~Gomis, and M.~de~Roo, ``{'Stringy' Newton-Cartan
  Gravity},'' {\em Class. Quant. Grav.}, vol.~29, p.~235020, 2012.

\bibitem{Bergshoeff:2018yvt}
E.~Bergshoeff, J.~Gomis, and Z.~Yan, ``{Nonrelativistic String Theory and
  T-Duality},'' {\em JHEP}, vol.~11, p.~133, 2018.

\bibitem{Bergshoeff:2018vfn}
E.~A. Bergshoeff, K.~T. Grosvenor, C.~Simsek, and Z.~Yan, ``{An Action for
  Extended String Newton-Cartan Gravity},'' {\em JHEP}, vol.~01, p.~178, 2019.

\bibitem{Bergshoeff:2017dqq}
E.~Bergshoeff, A.~Chatzistavrakidis, L.~Romano, and J.~Rosseel,
  ``{Newton-Cartan Gravity and Torsion},'' {\em JHEP}, vol.~10, p.~194, 2017.

\bibitem{Gomis:2000bd}
J.~Gomis and H.~Ooguri, ``{Nonrelativistic closed string theory},'' {\em J.
  Math. Phys.}, vol.~42, pp.~3127--3151, 2001.

\bibitem{Gomis:2019zyu}
J.~Gomis, J.~Oh, and Z.~Yan, ``{Nonrelativistic String Theory in Background
  Fields},'' 2019.

\bibitem{nullReduction}
C.~Duval, G.~Burdet, H.~P. K\"unzle, and M.~Perrin, ``Bargmann structures and
  newton-cartan theory,'' {\em Phys. Rev. D}, vol.~31, pp.~1841--1853, Apr
  1985.

\bibitem{YellowBook}
P.~Deligne and I.~Study, {\em Quantum Fields and Strings: A Course for
  Mathematicians}.
\newblock No.~v. 2 in Quantum Fields and Strings: A Course for Mathematicians,
  American Mathematical Soc., 1999.

\bibitem{connection2}
G.~Festuccia, D.~Hansen, J.~Hartong, and N.~A. Obers, ``Torsional newton-cartan
  geometry from the noether procedure,'' {\em Phys. Rev. D}, vol.~94,
  p.~105023, Nov 2016.

\bibitem{ActionPrinciple}
D.~Hansen, J.~Hartong, and N.~A. Obers, ``Action principle for newtonian
  gravity,'' {\em Phys. Rev. Lett.}, vol.~122, p.~061106, Feb 2019.

\end{thebibliography}

\end{document}